\documentclass[12pt,a4paper,twoside]{article}

\usepackage[dvips]{graphics}
\usepackage{graphicx}
\usepackage{cite}

\def\Journal#1#2#3#4{{#1} {\bf #2} (#3) #4}

\def\PLB{{\em Phys. Lett.}  B}
\def\PL{\em Phys. Lett.}
\def\PRL{\em Phys. Rev. Lett.}
\def\PRD{{\em Phys. Rev.} D}

\def\EPJC{{\em Eur. Phys. J.} C}

\def\CPC{\em Comp. Phys. Comm.}
\def\ZFP{{\em Zeit.\ Phys.} C }
\def\PRT{\em Phys. Rept.}

\def\etal{{\it et al.}}

\def\NPPS{\em Nucl.\ Phys.\ Proc.\ Suppl.}

\def\be{\begin{equation}}
\def\ee{\end{equation}}
\def\bea{\begin{eqnarray}}
\def\eea{\end{eqnarray}}

\newcommand{\Opal}{\mbox{\rm OPAL}}

\newcommand{\bbbar}     {\ensuremath{\mathrm{b\bar{b}}}}
\newcommand{\qqbar}     {\ensuremath{\mathrm{q\bar{q}}}}

\newcommand{\qqbargg}     {\ensuremath{\mathrm{q\bar{q}gg}}}
\newcommand{\qqbarqqbar}     {\ensuremath{\mathrm{q\bar{q}q\bar{q}}}}

\newcommand{\epem}              {\ensuremath{\mathrm{e^+e^-}}}

\newcommand{\as}                {\ensuremath{\alpha_\mathrm{S}}}
\newcommand{\asi}                {\ensuremath{\alpha_{\mathrm{S},i}}}
\newcommand{\asrs}                {\ensuremath{\alpha_\mathrm{S}(\sqrt{s})}}

\newcommand{\assq}    {\ensuremath{\alpha_\mathrm{S}^{\mathrm{2}}}}
\newcommand{\ascu}    {\ensuremath{\alpha_\mathrm{S}^{\mathrm{3}}}}

\newcommand{\asmz}              {\ensuremath{\alpha_\mathrm{S}(M_{\mathrm{Z^0}})}}

\newcommand{\zzero}     {\ensuremath{\mathrm{Z^0}}}

\newcommand{\mz}                {\ensuremath{M_{\mathrm{Z^0}}}}

\newcommand{\evis}              {\ensuremath{E_{\mathrm{vis}}}}

\newcommand{\pmiss}     {\ensuremath{p_{\mathrm{miss}}}}
\newcommand{\pbal}              {\ensuremath{p_{\mathrm{bal}}}}

\newcommand{\chisq}     {\ensuremath{\chi^2}}
\newcommand{\chisqd}    {\ensuremath{\chi^2/\mathrm{d.o.f.}}}
\newcommand{\xmu}               {\ensuremath{x_{\mu}}}

\newcommand{\xmumin}               {\ensuremath{x_{\mu}^\mathrm{min}}}
\newcommand{\xmuopt}               {\ensuremath{x_{\mu}^\mathrm{opt}}}

\newcommand{\ntrkl}             {\ensuremath{N_{\mathrm{long}}}}

\newcommand{\ycut}              {\ensuremath{y_{\mathrm{cut}}}}

\newcommand{\stat}              {\ensuremath{\mathrm{(stat.)}}}

\newcommand{\expt}               {\ensuremath{\mathrm{(exp.)}}}
\newcommand{\had}               {\ensuremath{\mathrm{(had.)}}}
\newcommand{\theo}              {\ensuremath{\mathrm{(theo.)}}}

\newcommand{\rs}                {\ensuremath{\sqrt{s}}}
\newcommand{\rsp}               {\ensuremath{\sqrt{s'}}}

\newcommand{\invpb}     {\ensuremath{\mathrm{pb}^{-1}}}

\newcommand{\py}                {PYTHIA}
\newcommand{\hw}                {HERWIG}
\newcommand{\ar}                {ARIADNE}
\newcommand{\jt}                {JETSET}

\newcommand{\debr}    {DEBRECEN 2.0}
\newcommand{\cdet}    {\ensuremath{C^{\mathrm{detector}}}}

\newcommand{\result} {\ensuremath{\asmz=0.1159\pm0.0004\stat\pm0.0012\expt\pm0.0024\had\pm0.0007\theo}}
\newcommand{\restot} {\ensuremath{\asmz=0.1159\pm0.0028~(\mathrm{total~error})}}
\newcommand{\resultxmu} {\ensuremath{\asmz=0.1161\pm0.0005\stat\pm0.0012\expt\pm0.0022\had}}
\newcommand{\resultxmin} {\ensuremath{\asmz=0.1155\pm0.0004\stat\pm0.0012\expt\pm0.0022\had}}
\newcommand{\resultthirty} {\ensuremath{\as(34.8\mathrm{GeV})=0.1368\pm0.0005\stat\pm0.0017\expt\pm0.0032\had\pm0.0011\theo}}

\newcommand{\Lep}{\mbox{LEP}}

\newcommand{\Pythia}{\mbox{PYTHIA}}

\begin{document}

\begin{titlepage}

\centerline{{\large Max-Planck-Institut f\"ur Physik}} \bigskip

\begin{flushright}
 MPP-2006-161 \\
\today
\end{flushright}

\bigskip\bigskip\bigskip

\begin{center}\textbf{
\Large Measurement of the Strong Coupling \boldmath{\as} from the 
Four-Jet Rate in \epem\ Annihilation using JADE data }
\end{center}

{\Large \par}

\bigskip

\begin{center}
  {\Large J. Schieck$^{a}$, S. Bethke$^{a}$, O. Biebel$^{b}$, S. Kluth$^{a}$, 
    P. A. Movilla Fern{\'a}ndez$^{c}$,  C. Pahl$^{a}$ and the 
    JADE Collaboration\footnote{See~\cite{naroska87} for
      the full list of authors}}
\end{center}

\par

\bigskip

\begin{abstract}
  Data from \epem\ annihilation into hadrons collected by the JADE
  experiment at centre-of-mass energies between 14~GeV and 44~GeV are
  used to study the four-jet event production rate as a function of the 
  Durham jet
  algorithm's resolution parameter \ycut. The four-jet rate is
  compared to QCD next-to-leading order calculations including resummation
  of large logarithms in the next-to-leading logarithmic approximation.
   The strong coupling measured from the
  four-jet rate is
\begin{center}
\result, \\
\end{center}
in agreement with the world average.\\
\end{abstract}
\bigskip\bigskip\bigskip\bigskip\bigskip
\bigskip
\bigskip
\vfill
$^{a}$ Max-Planck-Institut f\"ur Physik, F\"ohringer Ring 6, 
D-80805 M\"unchen, Germany \newline
$^{b}$ Ludwig-Maximilians-Universit\"at M\"unchen, Am Coulombwall 1, D-85748 Garching, Germany\newline
$^{c}$ Lawrence Berkeley National Laboratory, 1 Cyclotron Rd., Berkeley, CA 94720, U.S.A.\newline
\end{titlepage}

\section{Introduction}
The annihilation of electrons and positrons into hadrons allows
precise tests of Quantum Chromodynamics (QCD). Many observables
have been devised which provide a convenient way of characterizing
the main features of such events.
Multijet event rates are predicted in perturbation theory as functions
of the jet-resolution parameter, with one free parameter, the
strong coupling \as.
Events with four quarks in the final state, \qqbarqqbar, or two quarks and two 
gluons, \qqbargg, may lead to events with four-jet structure.
In leading order perturbation theory, the rate of four-jet events
in \epem\ annihilation is predicted to be proportional to \assq.
The strong coupling can be measured by determining the 
four-jet event production rate and fitting
the theoretical prediction to the data.\\
Calculations beyond leading order are made possible by theoretical
progress achieved during the last few years. 
For multi-jet rates as well as numerous event-shape 
distributions with perturbative expansions
starting at $\cal{O}(\as)$, matched next-to-leading order (NLO) 
and next-to-leading logarithmic approximations (NLLA) provide 
a satisfactory description of the
data over large kinematically allowed regions at many centre-of-mass
energies ~\cite{durham,dixon97,nagy98a,nagy98b}. \\
First evidence for four-jet structure has been reported earlier by
the JADE collaboration~\cite{bartel82b}. In addition multi-jet event 
production rates were measured and the three-jet rate was used to 
determine the the strong 
coupling \as~\cite{JadeMultiJet,JADEOPALJetR,JadeAlphaS}. 
The ALEPH, DELPHI and OPAL collaborations published measurements of \as\ based
on the four-jet rate
in the energy range between 91 and 209 GeV~\cite{aleph249,delphir4,OPALPN527}.
The same theoretical predictions as used here were employed to determine
the strong coupling \as. \\
In this analysis we use data collected by the JADE experiment in the
years 1979 to 1986 at the PETRA \epem\ collider at DESY at six
centre-of-mass energies spanning the range of 14--44~GeV. 
\section{Observable}
\label{theory}
Jet algorithms are applied to cluster the large number
of particles of a hadronic event into a small number of jets,
reflecting the parton structure of the event. For this
analysis we use the Durham scheme~\cite{durham}. Defining 
each particle initially to be a proto-jet, a resolution variable $y_{ij}$ 
is calculated for each pair of proto-jets $i$ and $j$:
\be
 y_{ij}=\frac{2\mathrm{min}(E_i^2,E_j^2)}{E_{\mathrm{vis}}^2}(1-\cos\theta_{ij}),
\ee
where $E_{i}$ and  $E_{j}$ are the energies of jets $i$ and $j$,
$\cos\theta_{ij}$ is the cosine of the angle between them 
and $E_{\mathrm{vis}}$ is the sum of the  energies
of the detected particles in the event (or the
partons in a theoretical calculation). 
If the smallest
value of $y_{ij}$ is less than a predefined value $\ycut$, the pair
is replaced by a new proto-jet with four-momentum
$p_{k}^\mu =  p_i^\mu + p_j^\mu$, and the clustering starts again.
Clustering ends
when the smallest value of $y_{ij}$ is larger than $\ycut$, and the remaining
proto-jets are accepted and counted as finally selected jets.\\
The next-to-leading order QCD calculation predicts the four-jet 
rate $R_{4}$, which is the fraction of four-jet events as a function
of \as. It can be written in the following way~\cite{nagy98b}:
\be
R_{4}(y_{\mathrm{cut}})=\frac{\sigma_{\mathrm{\mbox{\scriptsize{4-jet}}}}(y_{\mathrm{cut}})}{\sigma_{\mbox{\scriptsize tot}}} \\
   = \eta^{2}B_4(y_{\mathrm{cut}})+\eta^{3}[C_4(y_{\mathrm{cut}})+3/2(\beta_{0}\log{\xmu}
     -1)\ B_4(y_{\mathrm{cut}})]
\label{NLOcalc}        
\ee
where $\sigma_{\mathrm{\mbox{\scriptsize{4-jet}}}}(\ycut)$ is the cross-section
for the production of  hadronic events with four jets at fixed \ycut,
$\sigma_{\mathrm{\mbox{\scriptsize tot}}}$ the total hadronic cross-section,
$\eta = \as \ C_{\mathrm F} / 2\pi$ with $C_{\mathrm F}=4/3$ the colour
factor from the SU(3) symmetry group, $\xmu=\mu/\sqrt{s}$ with 
$\mu$ being the renormalization scale, $\sqrt{s}$ the centre-of-mass energy, 
and $\beta_{0}=(11-2n_{f}/3)$ with $n_{f}$ the number of active 
flavours\footnote{In this analysis the number of active flavours is set to five.}.  
The coefficients $B_4$ and $C_4$ are obtained by
integrating the massless matrix elements for \epem\
annihilation into four- and five-parton final states, calculated
by the program~\debr~\cite{nagy98b}.\\
Eq.~(\ref{NLOcalc}) is used to predict the four-jet rate
as a function of $y_{\mathrm{cut}}$. The fixed-order perturbative
prediction is not reliable for small values of $y_{\mathrm{cut}}$, due
to terms of the form $\as^{n}\ln^{m}(y_{\mathrm{cut}})$ which enhance the higher order 
corrections. An all-order resummation of such terms in the NLLA is possible 
for the Durham clustering algorithm ~\cite{durham}.
The NLLA calculation is combined with the NLO-prediction using the 
so-called modified R-matching scheme~\cite{OPALPR330}.
In the modified R-matching scheme the terms proportional to
$\eta^2$ and $\eta^3$ are removed from the NLLA 
prediction $R^{\mathrm{NLLA}}$ and the difference is then added to the NLO
calculation $R^{\mathrm{NLO}}$ with the result:
\be
 R^{\mathrm{R-match}}=R^{\mathrm{NLLA}}+[\eta^{2}(B_{4}-B^{\mathrm{NLLA}})+\eta^{3}(C_{4}-C^{\mathrm{NLLA}}-3/2(B_{4}-B^{\mathrm{NLLA}}))],
\label{NLLA}
\ee
where $B^{\mathrm{NLLA}}$ and $C^{\mathrm{NLLA}}$ are the coefficients of the 
expansion of $R^{\mathrm{NLLA}}$ as in Eq.~(\ref{NLOcalc}) and the
\xmu\ term and \ycut\ dependence have been suppressed for clarity. \\
\section{Analysis Procedure}
\subsection{The JADE Detector}
\label{sec_detector}
A detailed description of the JADE detector can be found
in~\cite{naroska87}. This analysis relies mainly on the reconstruction
of charged particle trajectories and on the measurement of energy
deposited in the electromagnetic calorimeter.  Tracking of charged
particles was performed with the central tracking detector, which was
positioned in a solenoidal magnet providing an axial magnetic field of
0.48 T. The central detector contained a large volume jet chamber.
Later a vertex chamber close to the interaction point and surrounding
$z$-chambers to measure the $z$-coordinate~\footnote{In the JADE
right-handed coordinate system the $+x$ axis points towards the
centre of the PETRA ring, the $y$ axis pointed upwards and the $z$
axis points in the direction of the positron beam. The polar angle
$\theta$ and the azimuthal angle $\phi$ are defined with respect to
$z$ and $x$, respectively, while $r$ is the distance from the
$z$-axis.} were added. Most of the tracking information is obtained
from the jet chamber, which provides up to 48 measured space points
per track, and good tracking efficiency in the region $|\cos
\theta|<0.97$. Electromagnetic showers are measured
by the lead glass
calorimeter surrounding the magnet coil, separated into a barrel
($|\cos \theta|<0.839$) and two end-cap ($0.86<|\cos \theta|<0.97$)
sections.  The electromagnetic calorimeter consisted of 2520 lead
glass blocks with a depth of 12.5 radiation lengths in the barrel
(since 1983 increased to 15.7 in the middle 20\% of the barrel) and
192 lead glass blocks with 9.6 radiation lengths in the end-caps.
\subsection{Data Samples}
The data used in this analysis were collected by JADE between 1979 and
1986 and correspond to a total integrated luminosity of 195 \invpb.
The breakdown of the data samples, average centre-of-mass energy, energy
range, data taking period, collected integrated luminosities and the 
size of the
data samples after selection of hadronic events are given in
table~\ref{lumi}. 
The data samples are chosen following previous analyses,
e.g.~\cite{naroska87,jadenewas,JADEOPALJetR,movilla02b,pedrophd}~\footnote{The
data are available from two versions of the reconstruction software
from 9/87 and from 5/88.  We use the set from 5/88 as the default version
and consider differences
between the results as an experimental systematic uncertainty.}.
\begin{table}[htb!]
\begin{center}
\begin{tabular}{|r|r|r|r|r|} \hline
average       & energy       & year & luminosity  &  selected  \\
energy in GeV & range in GeV &      & (\invpb)    &  events  \\
\hline
14.0 & 13.0--15.0 & 1981       & 1.46 &  1783 \\
\hline
22.0 & 21.0--23.0 & 1981       & 2.41 &  1403 \\
\hline
34.6 & 33.8--36.0 & 1981--1982 & 61.7 &  14313 \\
35.0 & 34.0--36.0 & 1986       & 92.3 &  20876 \\
\hline
38.3 & 37.3--39.3 & 1985       & 8.28 &  1585 \\
\hline
43.8 & 43.4--46.4 & 1984--1985 & 28.8 &  4374 \\
\hline
\end{tabular}
\end{center}
\caption{
The average centre-of-mass energy, the energy range, the year
of data taking and the integrated luminosity for each data
sample, together with the numbers of selected data events.}
\label{lumi}
\end{table}
\subsection{Monte Carlo Samples}
Samples of Monte Carlo simulated events are used to correct the data
for experimental \linebreak acceptance, resolution and 
backgrounds. The process
$\epem\to\mathrm{hadrons}$ is simulated using  \linebreak \py~5.7~\cite{jetset3}.
Corresponding samples using \hw~5.9~\cite{herwig} are used
for systematic checks.  The Monte Carlo samples generated at each
energy point are processed through a full simulation of the
JADE detector~\cite{jadesim1,jadesim2,jadesim3,jadesim4}, and reconstructed in
essentially the same way as the data. 
In addition, for comparisons
with the corrected data, and when correcting for the effects of
hadronization, large samples of Monte Carlo events without detector
simulation are employed, using the parton shower models \py~6.158,
\hw~6.2 and \ar~4.11~\cite{ariadne3}.  
Each of these hadronization
models contains a number of tunable parameters; 
these were adjusted by tuning to previously published \Opal\ data at 
$\sqrt{s}\sim91$~GeV as summarized in ~\cite{OPALPR141} for 
\py/\jt\ and in ~\cite{OPALPR379} for \hw\ and \ar. 
The data taken by the JADE experiment are well described 
by the Monte Carlo simulated events~\cite{pedrophd}, with only the
centre-of-mass energy changed and set to the respective value
of the event.
\subsection{Selection of Events}
The selection of events for this analysis aims to identify hadronic
event candidates and to reject events with a large amount of energy
emitted by initial state photon radiation (ISR).  The selection of hadronic
events is based on cuts on event multiplicity to remove leptonic
final states and on visible energy and longitudinal momentum imbalance
to remove radiative and two-photon events, $\epem \to \epem +$
hadrons.  The cuts are documented 
in~\cite{JadeAlphaS,StdEvSel1,StdEvSel2} and summarized in 
a previous publication~\cite{jadenewas}:\\
Standard criteria are used to select good tracks and clusters of
energy deposits in the calorimeter for subsequent analysis.  
For charged particle tracks the pion mass was assigned while
for clusters zero mass particles are assumed.
Charged particle tracks are required to have at least 20
hits in $r-\phi$ and at least 12 hits in $r-z$ in the jet chamber.  The
total momentum is required to be at least 100~MeV.  Furthermore, the
point of closest approach of the track to the collision axis is
required to be less than 5~cm from the nominal collision point in the
$x-y$ plane and less than 35~cm in the $z-$direction. \\
In order to mitigate the effects of double counting of energy from
tracks and calorimeter clusters a standard algorithm is
adopted which associates charged particles
with calorimeter clusters, and subtracts the estimated contribution
of the charged particles from the cluster energy. 
Clusters in the electromagnetic calorimeter are required to 
have an energy exceeding
0.15~GeV after the subtraction of the expected energy deposit of 
associated tracks.
From all accepted tracks and clusters the visible energy
$\evis=\sum_i E_i$, momentum balance $\pbal=|\sum_i p_{z,i}|/\evis$ and
missing momentum $\pmiss=|\sum_i \vec{p}_i|$ are calculated.  \\
Hadronic event candidates are required to pass the following selection 
criteria:
\begin{itemize}
\item The total energy deposited in the electromagnetic calorimeter
  has to exceed 1.2~GeV (0.2~GeV) for $\rs<16$~GeV, 2.0~GeV (0.4~GeV)
  for $16<\rs<24$~GeV and 3.0~GeV (0.4~GeV) for $\rs>24$~GeV in the
  barrel (each endcap) of the detector.
\item Events with exactly four tracks with configurations where three
  tracks are in one hemisphere and one track is in the opposite hemisphere
  are rejected.
\item At least three tracks have to have more than 24 hits in $r-\phi$
  and a momentum in the $x-y$ plane larger than 500~MeV; 
  these tracks are called long tracks.
\item The visible energy has to fulfill $\evis/\rs>0.5$.
\item The momentum balance has to fulfill $\pbal<0.4$.
\item The missing momentum has to fulfill $\pmiss/\rs<0.3$.
\item The z-coordinate of the reconstructed event vertex has to lie
  within 15~cm of the interaction point.
\item The polar angle of the event thrust axis~\cite{thrust} is
  required to satisfy $|\cos(\theta_{\mathrm T})|<0.8$ in order that
  the events be well contained within the detector acceptance.
\item The number of good charged particle tracks is required to be
  greater than three reducing $\tau^{+}\tau^{-}$ and two-photon
  backgrounds to a negligible level.
\end{itemize}
The numbers of selected events for each centre-of-mass energy are shown in
table~\ref{lumi}.
\subsection{Corrections to the data}
\label{detectorcorrection}
All selected tracks and electromagnetic calorimeter clusters
are used in the evaluation of the four-jet rate.  
The four-jet rate distribution as a function of the jet resolution
\ycut\ after all selection cuts applied is called
the detector-level distribution.\\
In this analysis events from the process $\epem\rightarrow\bbbar$ 
systematically bias our results, since especially at low centre-of-mass energies
the large mass of the b quarks and of the subsequently produced and decaying 
B hadrons will influence the four-jet rate 
distribution~\cite{aleph249,delphir4,OPALPN527}. The QCD predictions are
calculated for massless quarks  and thus we choose to correct our data
for the presence of \bbbar\ events. About $1/11$ of all \qqbar\ events
are \bbbar\ events. The expected number of \bbbar\ background events $\eta_i$
is subtracted from the observed number of data events $N_i$ at each
\ycut\ bin $i$.  The effects of detector acceptance and resolution
and of residual ISR are then accounted for by a multiplicative
correction procedure. \\ 
Two distributions are formed from Monte Carlo 
simulated signal events; the first, at the detector-level,
treats the Monte Carlo events identically to the data, while the
second, at the hadron-level, is computed using the true four-momenta
of the stable particles\footnote{ All charged and 
neutral particles with a lifetime longer
than $3\times 10^{-10}$~s are treated as stable.}, 
and is restricted to events where $\rsp$, the centre-of-mass
energy reduced due to ISR, satisfies $\rs-\rsp<0.15$~GeV. 
The ratio of the Monte Carlo distribution taken at the hadron-level 
to the distribution taken at the detector-level for each \ycut\ bin
$i$, $\cdet_i$, is used as a correction factor for the data.  This
finally yields the corrected number of four jet events at \ycut\ bin
$i$, $\tilde{N_{i}}=\cdet_i\cdot(N_{i}-\eta_{i})$.  The hadron-level
distribution is then normalized at each \ycut\ bin $i$ by
calculating $R_{4,i}=\tilde{N_{i}}/N$, 
where the $N$ is the corrected number of events selected at hadron-level.
We observe some disagreement
between the detector corrections calculated using \py\ or \hw\ at low
centre-of-mass energies while at larger \rs\ the correction 
factors agree well within the
regions chosen for comparison with the theory predictions.
The difference in detector corrections is evaluated as an
experimental systematic uncertainty. 
The numerical results of the four-jet rate at hadron-level at the different
energy points are summarized in Tables~\ref{hadron_tab_1} 
and ~\ref{hadron_tab_2}. \\
\section{Systematic Uncertainties}
\label{systematic}
Several sources of possible systematic uncertainties are studied.
Uncertainties originating from massless quark calculations are not considered,
since contributions to the four-jet rate from B hadrons are 
subtracted at detector-level. For each variation of parameters
the difference of the resulting value of \as\  with respect to the default value
is taken as a systematic uncertainty.
The default value of \as\ is determined with the standard event selection 
and the correction procedure using \Pythia.
The systematic uncertainty is taken to be symmetric around
the default value of \as.
\subsection{Experimental uncertainties:}
Contributions to the experimental uncertainties are estimated by 
repeating the analysis with varied cuts or procedures.  For each systematic
variation the value of \as\ is determined and then compared to the
result of the standard analysis (default value).
\begin{enumerate}
\item In the standard analysis the reconstruction software from 5/88 is used.
  As a variation a different reconstruction software from 9/87 is used.
\item In the default method the estimated minimum ionizing energy 
  from tracks associated with electromagnetic calorimeter clusters is 
  subtracted from the cluster energies. As a variation all accepted tracks
  and all uncorrected electromagnetic clusters are used.
\item The thrust axis is required to satisfy $|\cos(\theta_{\mathrm
    T})| < 0.7$.  With this more stringent cut events are restricted
  to the barrel region of the detector, which provides better
  measurements of tracks and clusters compared to the endcap regions.
\item Instead of using \py\ for the correction of detector
  effects as described in section~\ref{detectorcorrection}, events
  generated with \hw\ are used.
\item The requirement on missing momentum is dropped or tightened to
  $\pmiss/\rs<0.25$. The larger deviation from the default value is taken as
a systematic uncertainty.
\item The requirement on the momentum balance is dropped or tightened
  to $\pbal<0.3$. The larger deviation from the default value is taken as
a systematic uncertainty.
\item The requirement on the number of long tracks is tightened to
  $\ntrkl\ge 4$.
\item The requirement on the visible energy is varied to $\evis/\rs>0.45$ and
  $\evis/\rs>0.55$. The larger deviation from the default value is taken as
a systematic uncertainty.
\item The fit range is changed. 
Two different cases are considered. First  the fit range 
is reduced by one data point at each edge of the standard fit range. 
Second the fit range is extended by one data point at each
edge of the standard fit range. The larger deviation from the default fit is 
taken as a systematic uncertainty. 
In order to take statistical fluctuations into account, 
the deviation is calculated using the average deviation of a 
fit applied to 50 Monte Carlo samples.
\item The amount of subtracted \bbbar\ background is varied
  by $\pm$5\% of its nominal value of about $1/11$ to 
  cover uncertainties in the estimation of
  the background fraction in the data. The larger deviation from 
  the default value is taken as the systematic uncertainty.
\end{enumerate}
All contributions listed above are added in quadrature and the result is 
quoted as the experimental systematic uncertainty. The dominating
effects are the use of the different data versions and the different
correction for detector effects.
\subsection{Hadronization:}
  The uncertainties associated with the hadronization correction
  (see section~\ref{fitprocedure}) are assessed by using \hw\ and
  \ar\ instead of the default hadronization correction using \py. 
  The larger change in \as\ resulting from these
  alternatives is taken to define the symmetric hadronization systematic
  uncertainty.
\subsection{Theoretical Uncertainties:}
The theoretical uncertainty, associated with missing higher 
order terms in the theoretical prediction, is assessed by varying 
the renormalization scale factor \xmu. The predictions of a
complete QCD calculation would be independent of \xmu, but a
finite-order calculation such as that used here retains some 
dependence on \xmu. The renormalization scale factor \xmu\ is set to 0.5 
and one. The larger 
deviation from the default 
value of \as\ is taken as systematic uncertainty.
\section{Results}
\subsection{Four-Jet Rate Distributions}
\begin{figure}[htb!]
\begin{tabular}[tbp]{cc}
\includegraphics[width=0.5\textwidth]{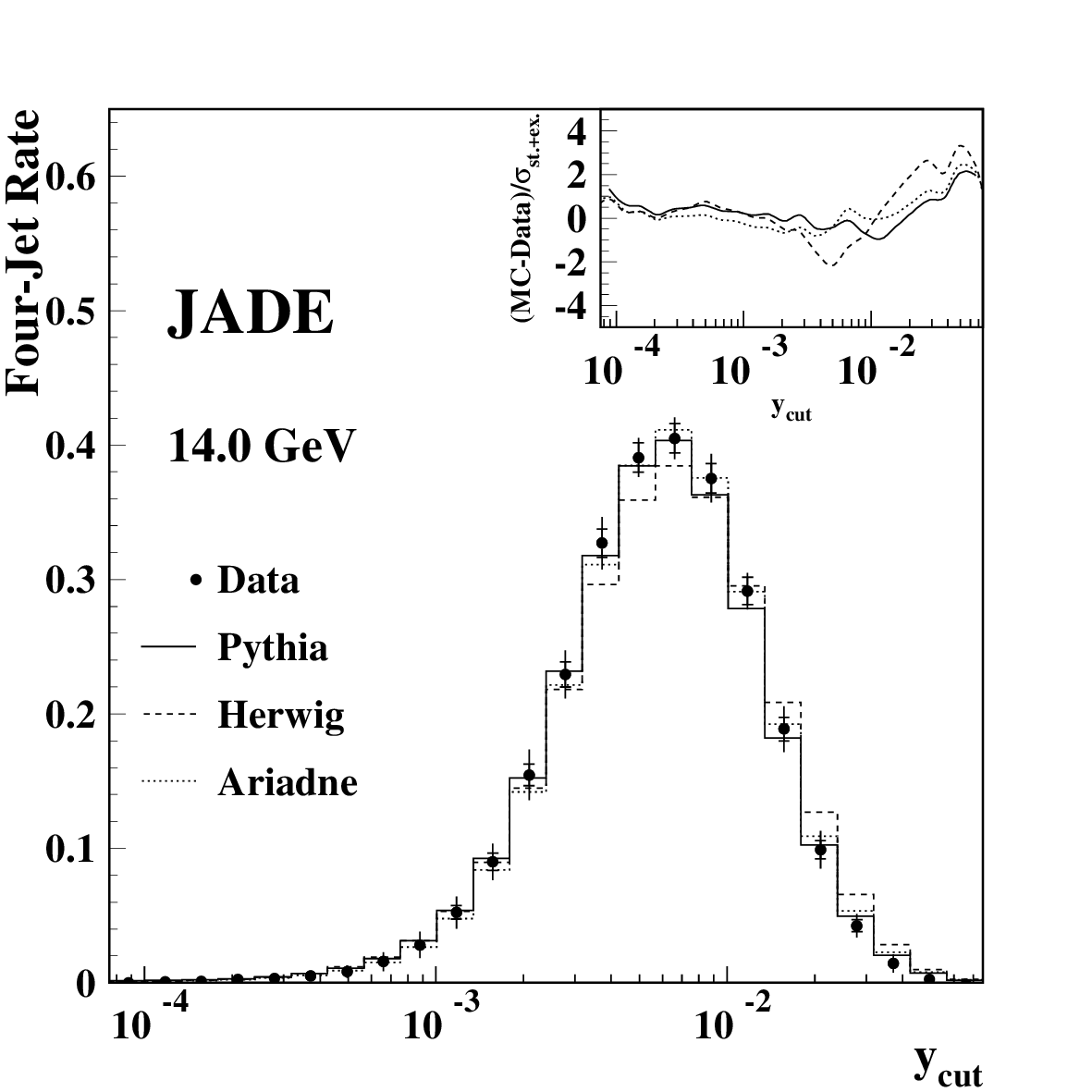} &
\includegraphics[width=0.5\textwidth]{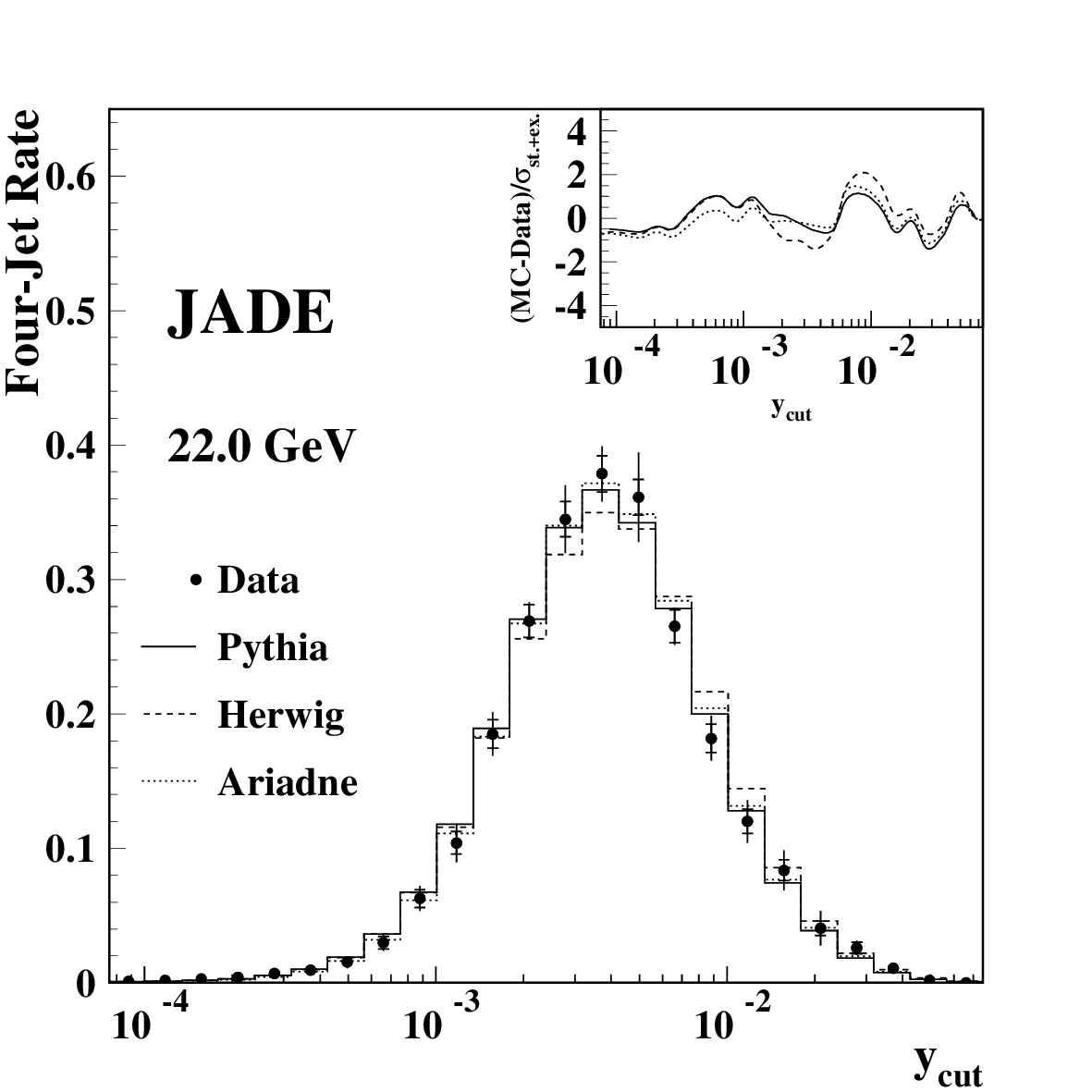} \\
\includegraphics[width=0.5\textwidth]{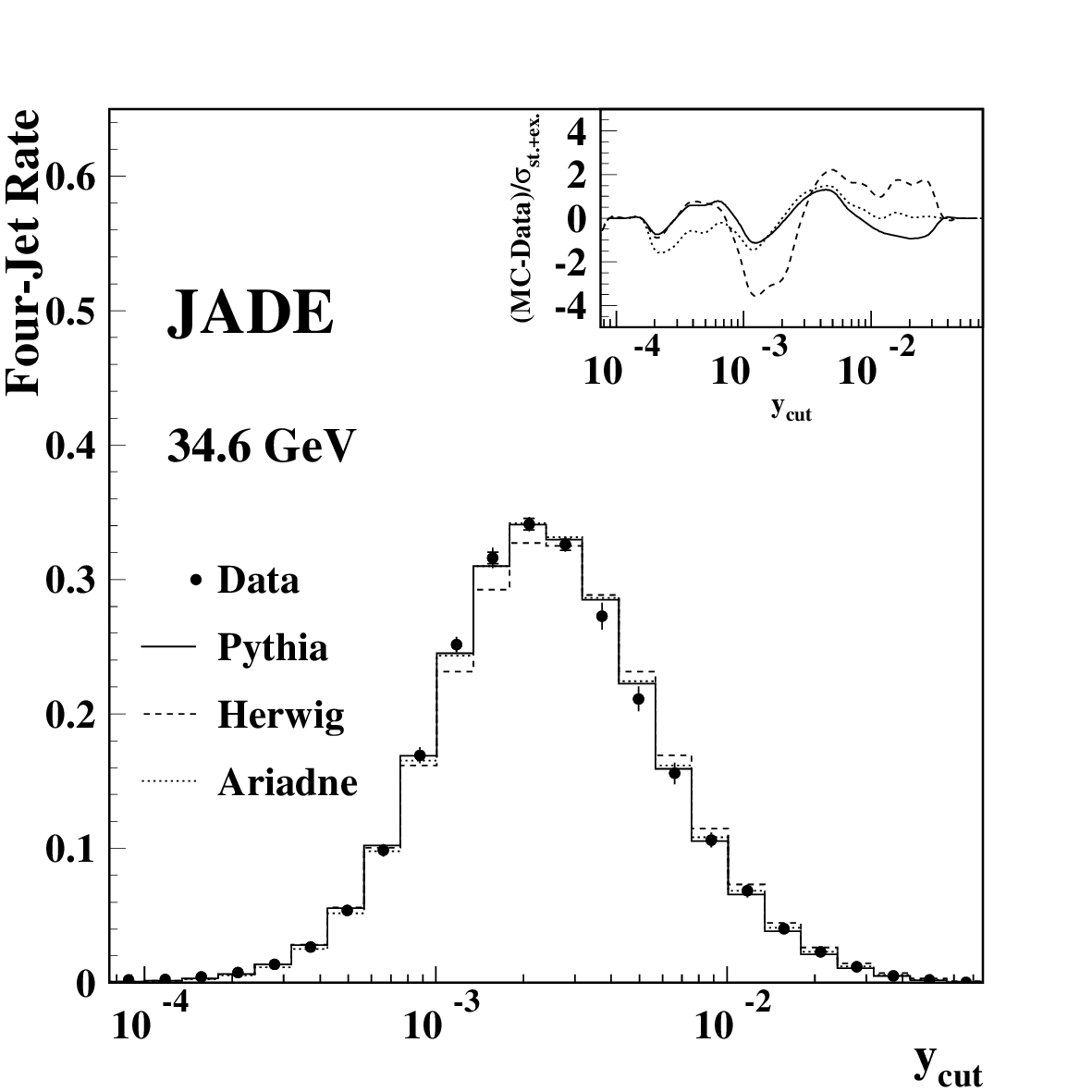} &
\includegraphics[width=0.5\textwidth]{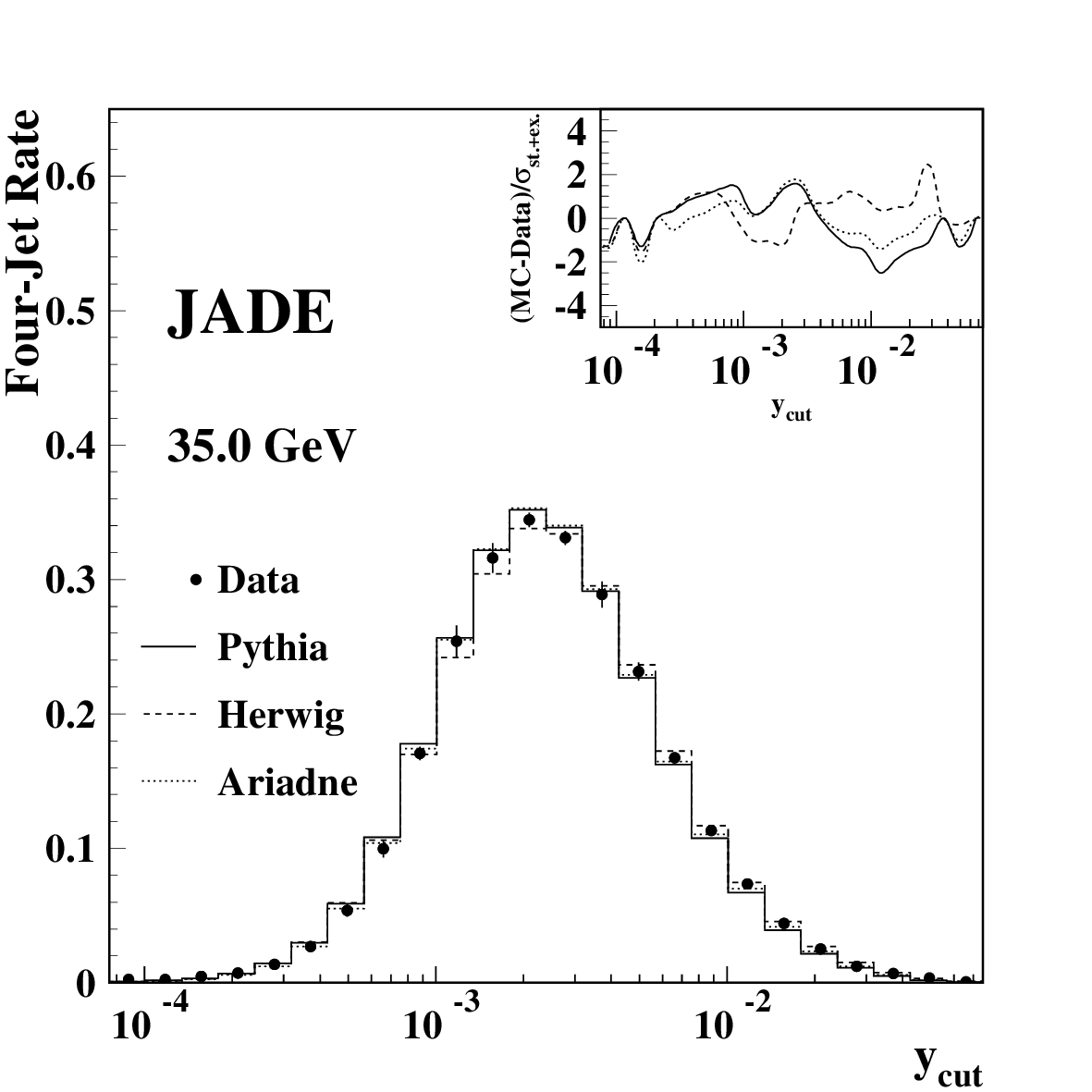} \\
\end{tabular}
\caption{The figures show the four-jet rate distribution corrected for 
  detector effects
  at hadron-level as a function of the \ycut\ resolution
  parameter obtained with the Durham algorithm.  The four-jet rate at
  four average centre-of-mass energies 
  are shown for $\rs=14$ to 35~GeV in comparison with
  predictions based on \py, \hw\ and \ar\ Monte Carlo events. The
  expected \bbbar\ background is subtracted from the data.
  The error bars show the statistical (inner part) and the experimental
 and statistical uncertainties added in quadrature. 
  The panel in each upper right corner shows the differences between
  data and Monte Carlo, divided by the quadratic sum of the statistical and
  experimental error. At points with no data events, the difference is
  set to zero. }
\label{hadron}
\end{figure}

\begin{figure}[htb!]
\begin{tabular}[tbp]{cc}
\includegraphics[width=0.5\textwidth]{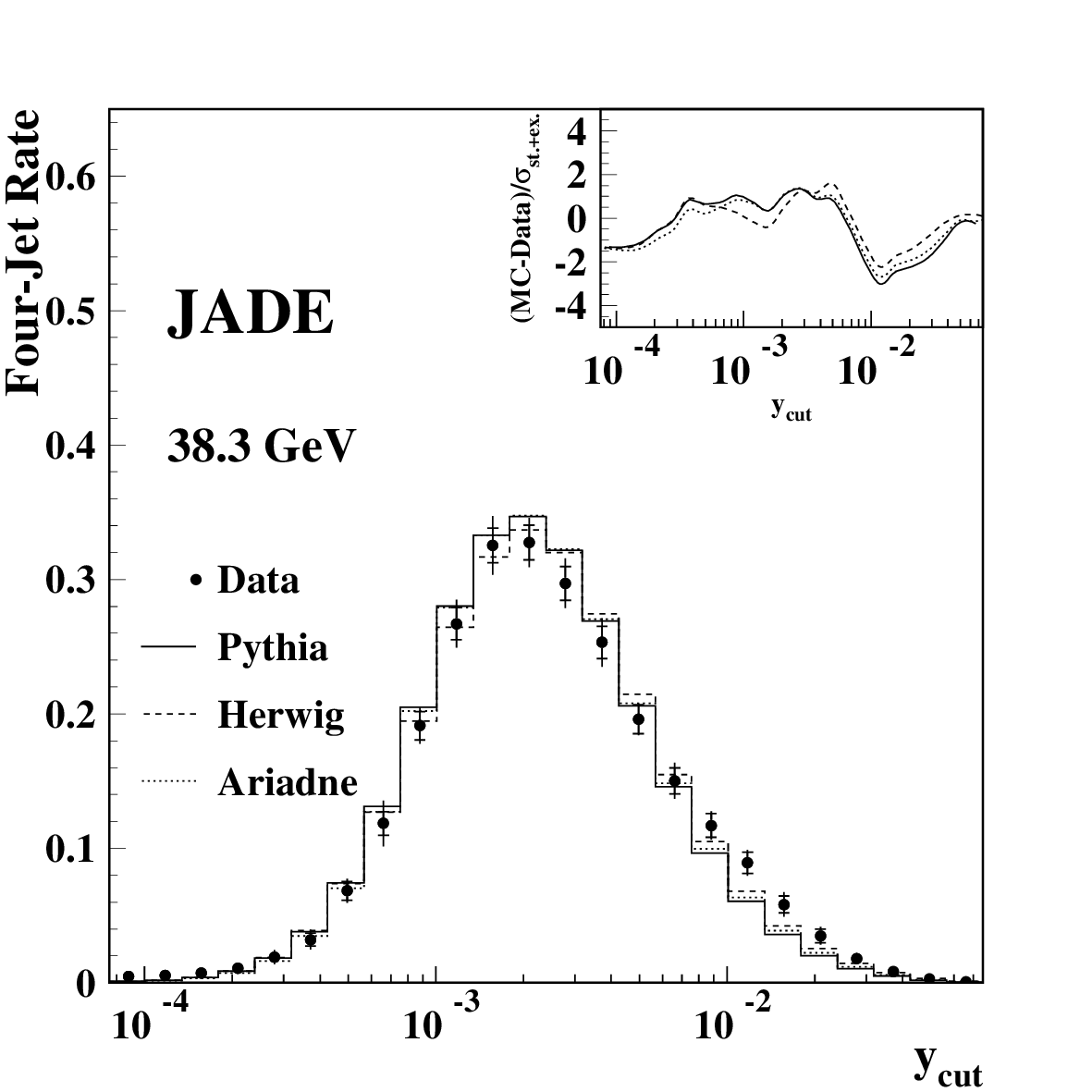} &
\includegraphics[width=0.5\textwidth]{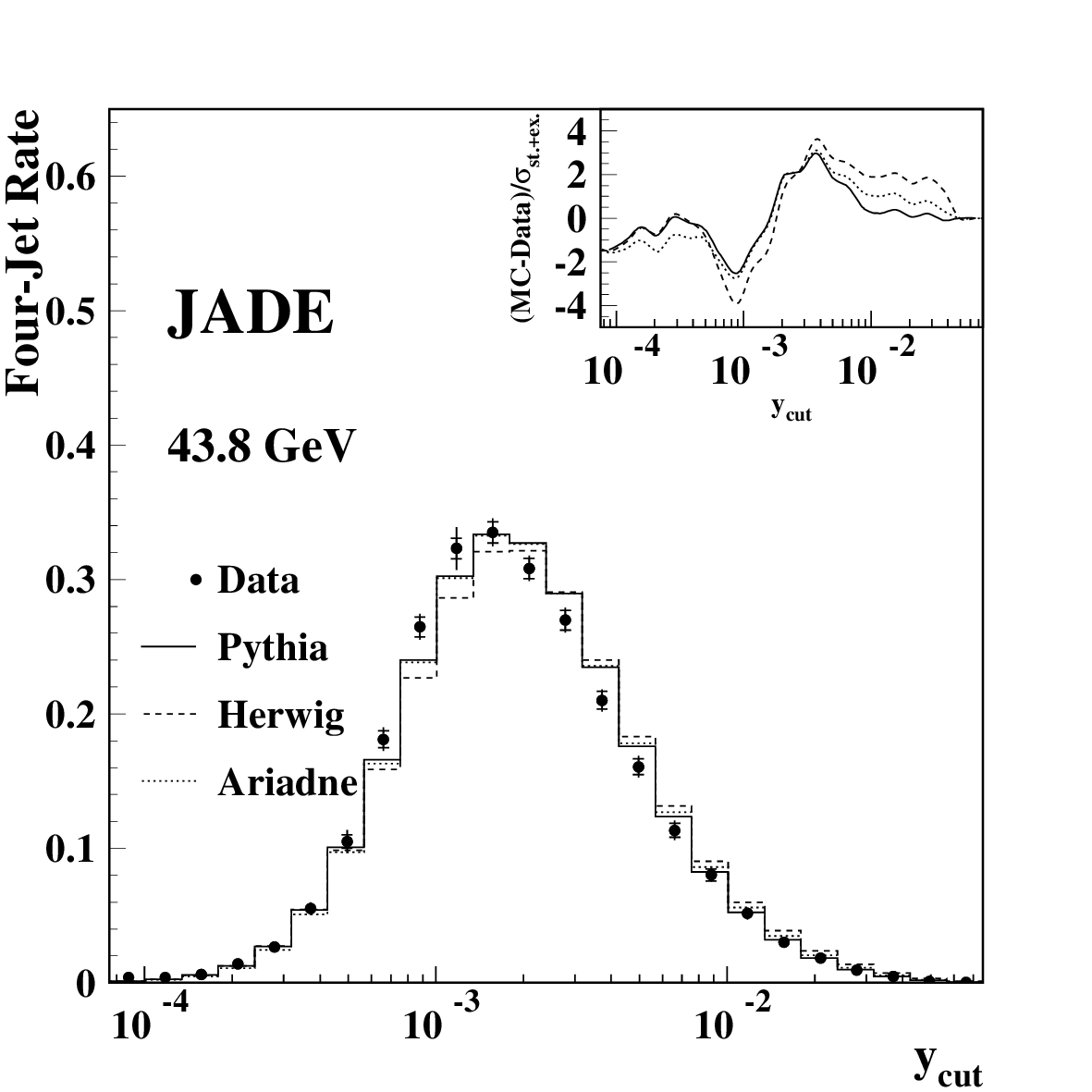} \\
\end{tabular}
\caption{ Same as figure~\ref{hadron} for $\rs=38.3$ and 43.8~GeV. }
\label{hadron2}
\end{figure}

The four-jet rates for the six centre-of-mass energy points
after subtraction of \bbbar\ background and correction for detector effects are shown in
figures~\ref{hadron} and~\ref{hadron2}.  Superimposed are the
distributions predicted by the \py, \hw\ and \ar\ Monte Carlo models.
Towards large \ycut\ values (right to the maximum of the distribution) the
decrease of the four-jet rate corresponds to the migration and
classification to three- and two-jet events. Towards smaller \ycut\
values (left to the maximum of the distribution) the decrease
corresponds to the migration and classification to five or more
jet events, i.e. towards the higher order QCD and non-perturbative or 
hadronization region.
In order to make a more clear comparison between data and models, the
inserts in the upper right corner show the differences between data
and each model, divided by the combined statistical and experimental
error at that point. The sum of squares of these differences would, in
the absence of point-to-point correlations, represent a $\chi^{2}$
between data and the model.  However, since correlations are present,
such $\chi^{2}$ values should be regarded only as a rough indication
of the agreement between data and the models. The three models are
seen to describe the data well.
\subsection{Determination of \boldmath{\as}}
\label{fitprocedure}
Our measurement of the strong coupling \as\ is based on
\chisq\ fits of QCD predictions to the corrected four-jet rate
distribution, i.e. the data shown in figures~\ref{hadron}
and~\ref{hadron2}.  The theoretical predictions of the four-jet rate
using the combined $\cal{O}(\ascu)$+NLLA calculation as described in
section~\ref{theory} provide distributions at the parton-level. 
In the Monte Carlo simulation the parton-level distributions are 
obtained from the partons after the parton shower has stopped, 
just before the hadronization. 
In order to confront the theory with the hadron-level data, it is
necessary to correct for hadronization effects.
This is done by calculating the
distributions at both the hadron and the parton-level
using \py\ and, as a cross-check, with the \hw\ and \ar\ models. 
The theoretical prediction is then multiplied by the ratio of the 
hadron- and parton-level distributions.\par
\begin{figure}[htb!]
\begin{tabular}[tbp]{cc}
\includegraphics[width=0.5\textwidth]{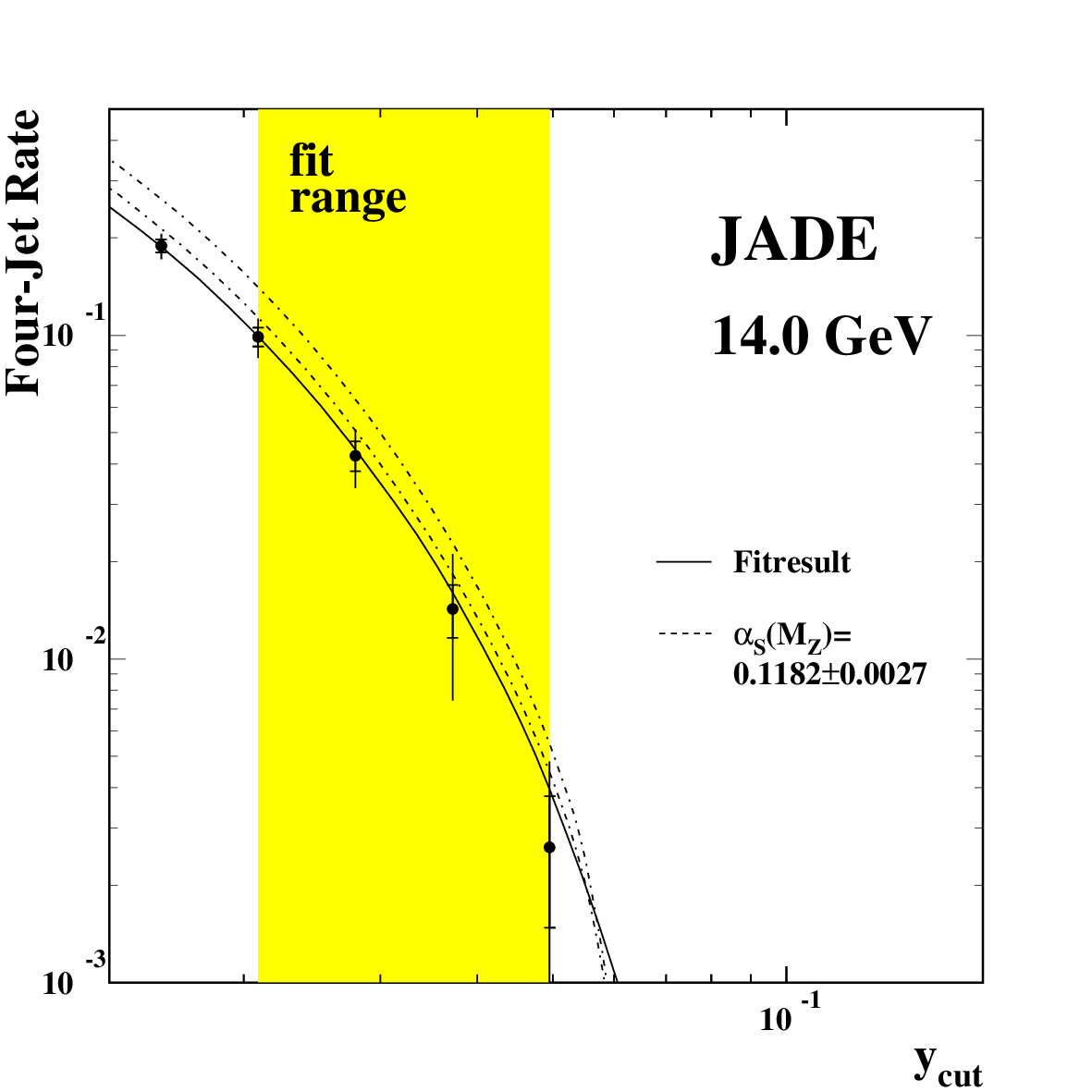} &
\includegraphics[width=0.5\textwidth]{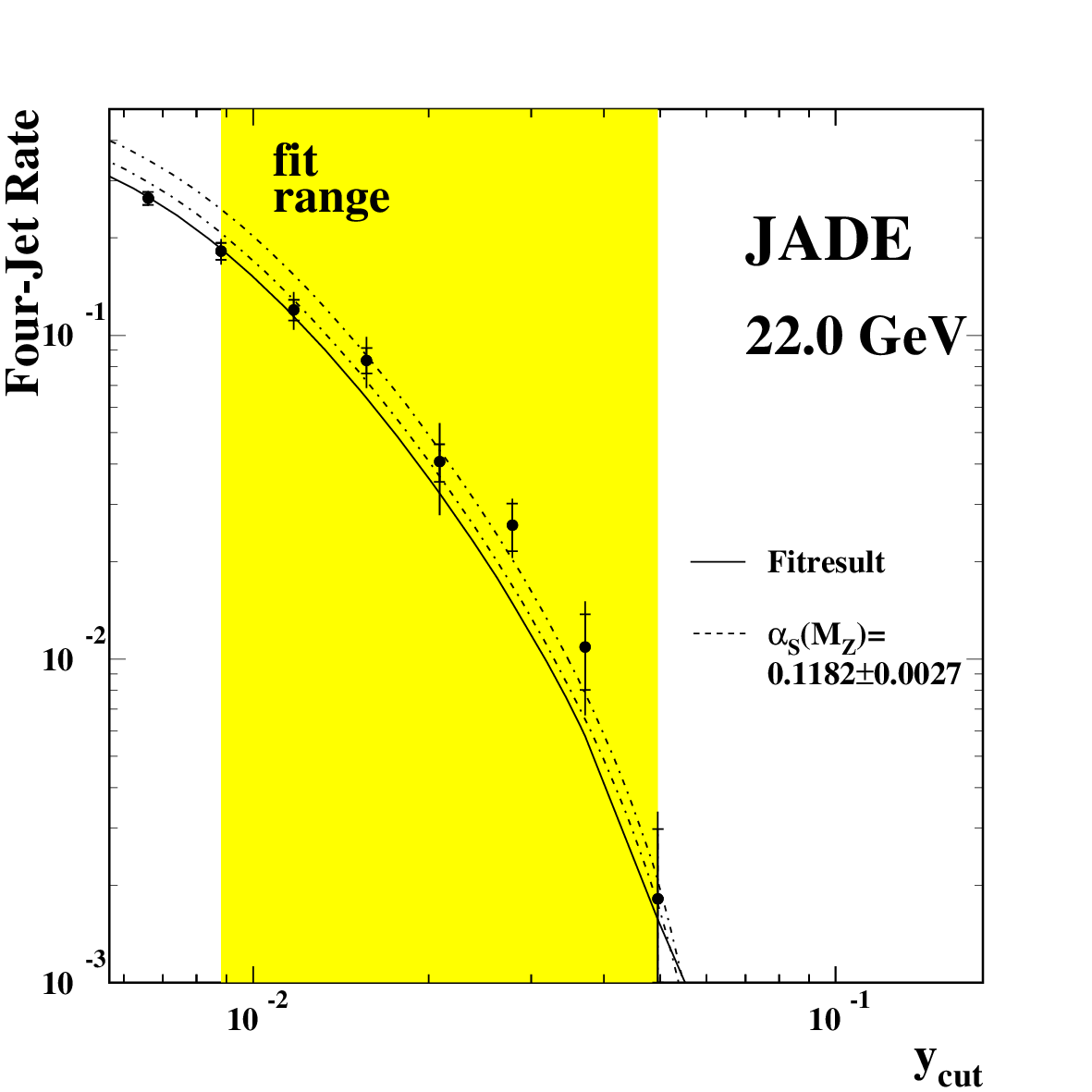} \\
\end{tabular}
\caption{ The plots show the four-jet rate distributions at the hadron-level
  for $\rs=14$~GeV and 22~GeV. 
  The error bars show the statistical (inner part) and the experimental
 and statistical uncertainties added in quadrature. 
The solid curve shows the theory prediction after $\chi^{2}$ minimization 
within the fit range indicated. 
The dash-dotted lines shows the error band of the four-jet rate prediction
with \asmz\ being the current world average value and its error~\cite{bethke04}.}
\label{fit_plot}
\end{figure}
\begin{figure}[htb!]
\begin{tabular}[tbp]{cc}
\includegraphics[width=0.5\textwidth]{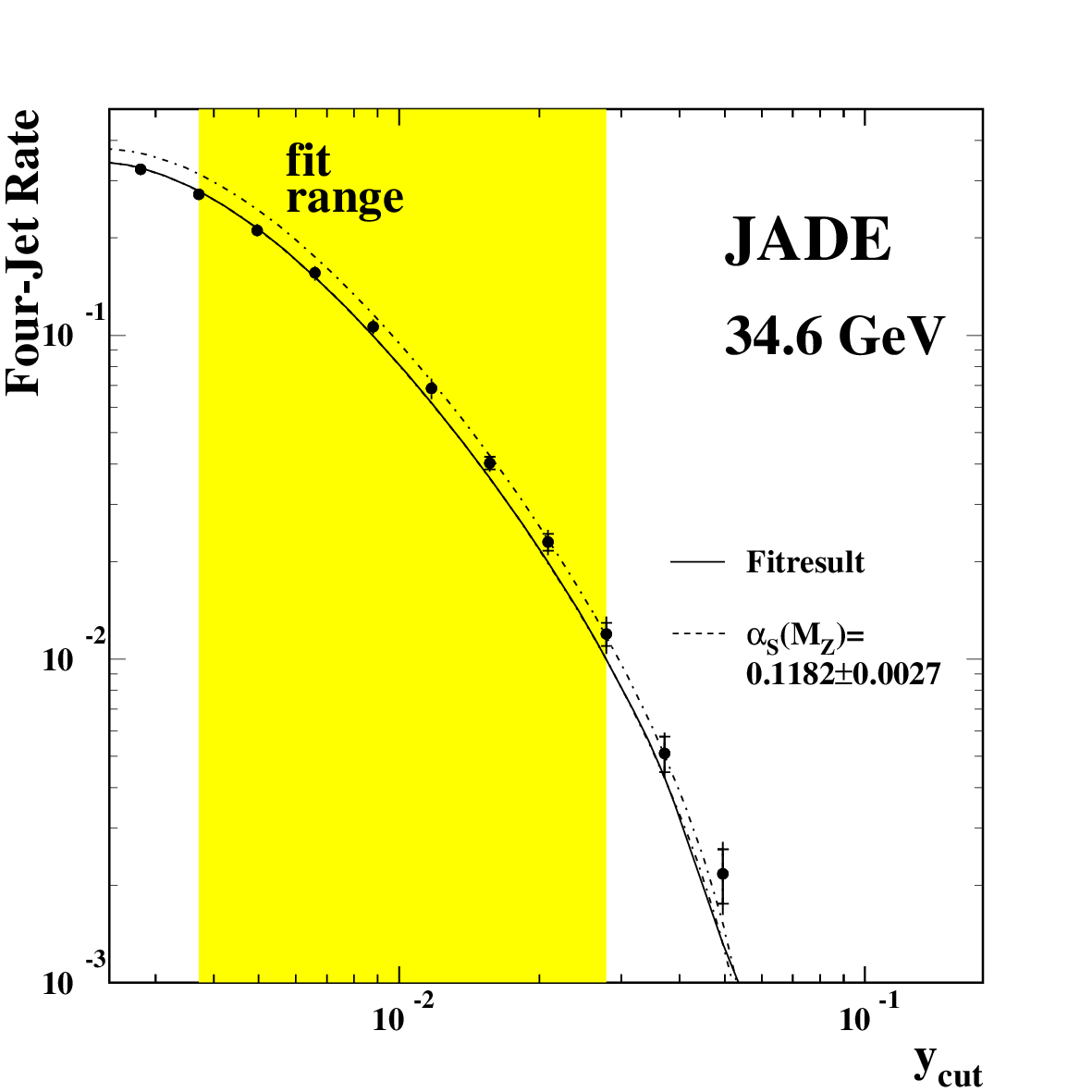} &
\includegraphics[width=0.5\textwidth]{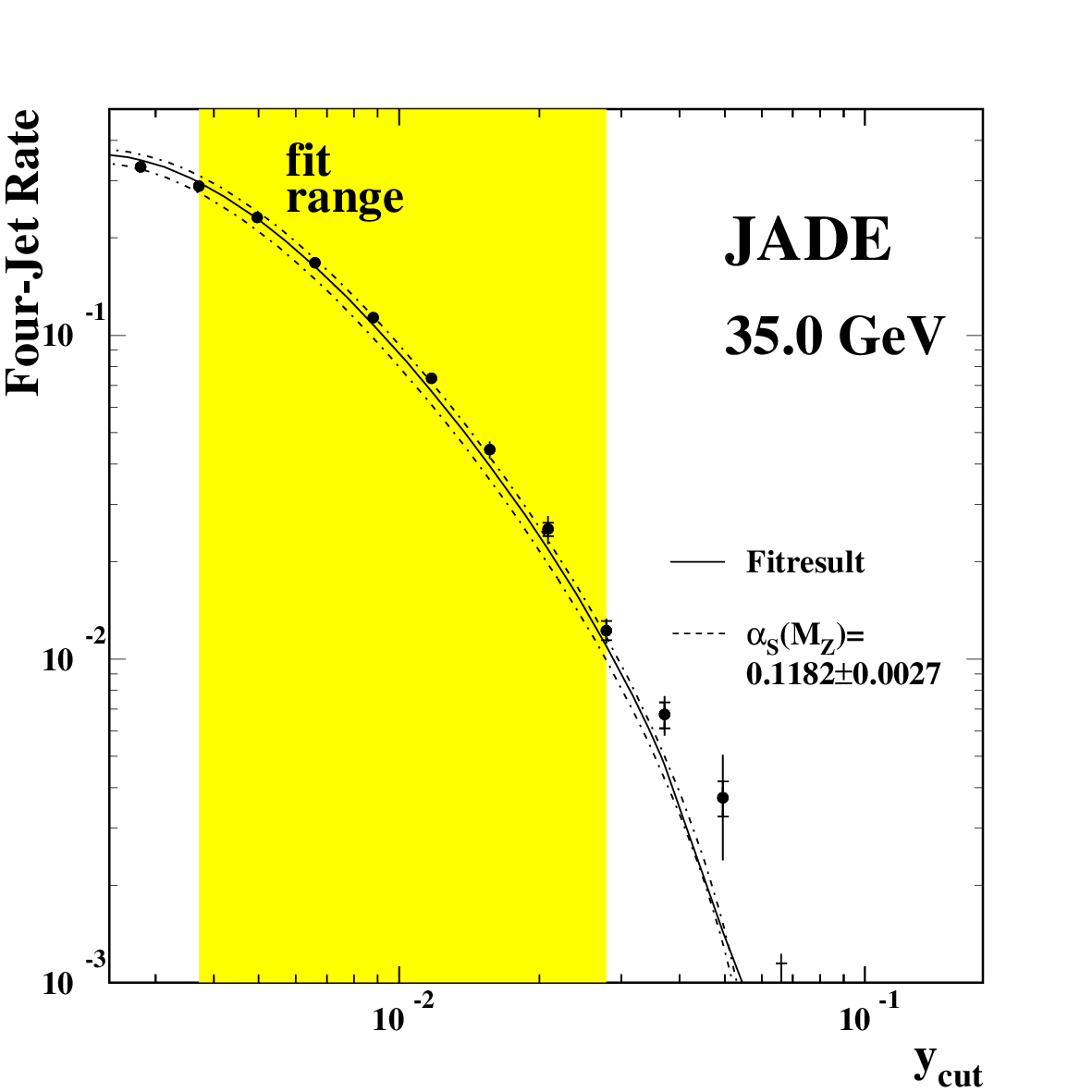} \\
\includegraphics[width=0.5\textwidth]{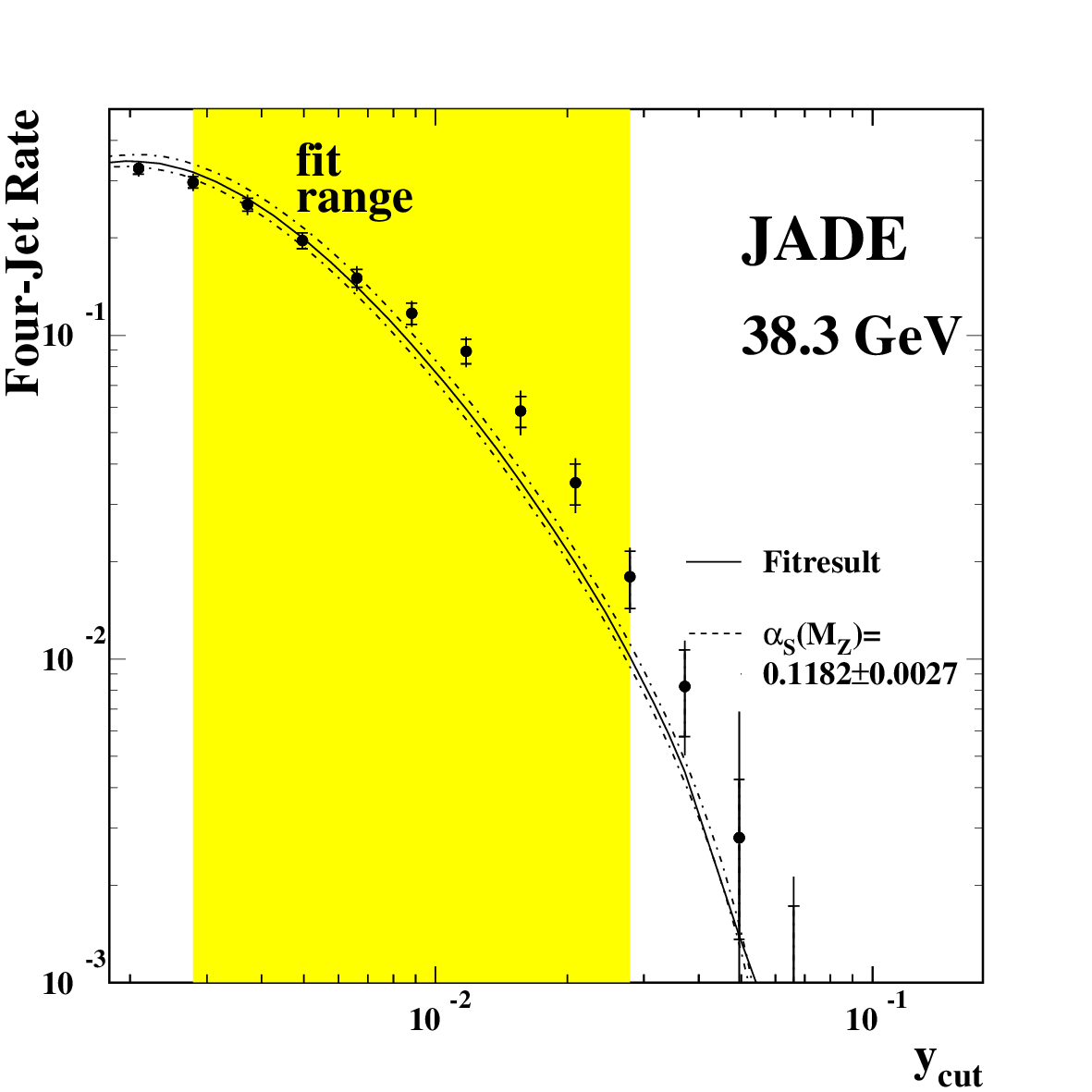} &
\includegraphics[width=0.5\textwidth]{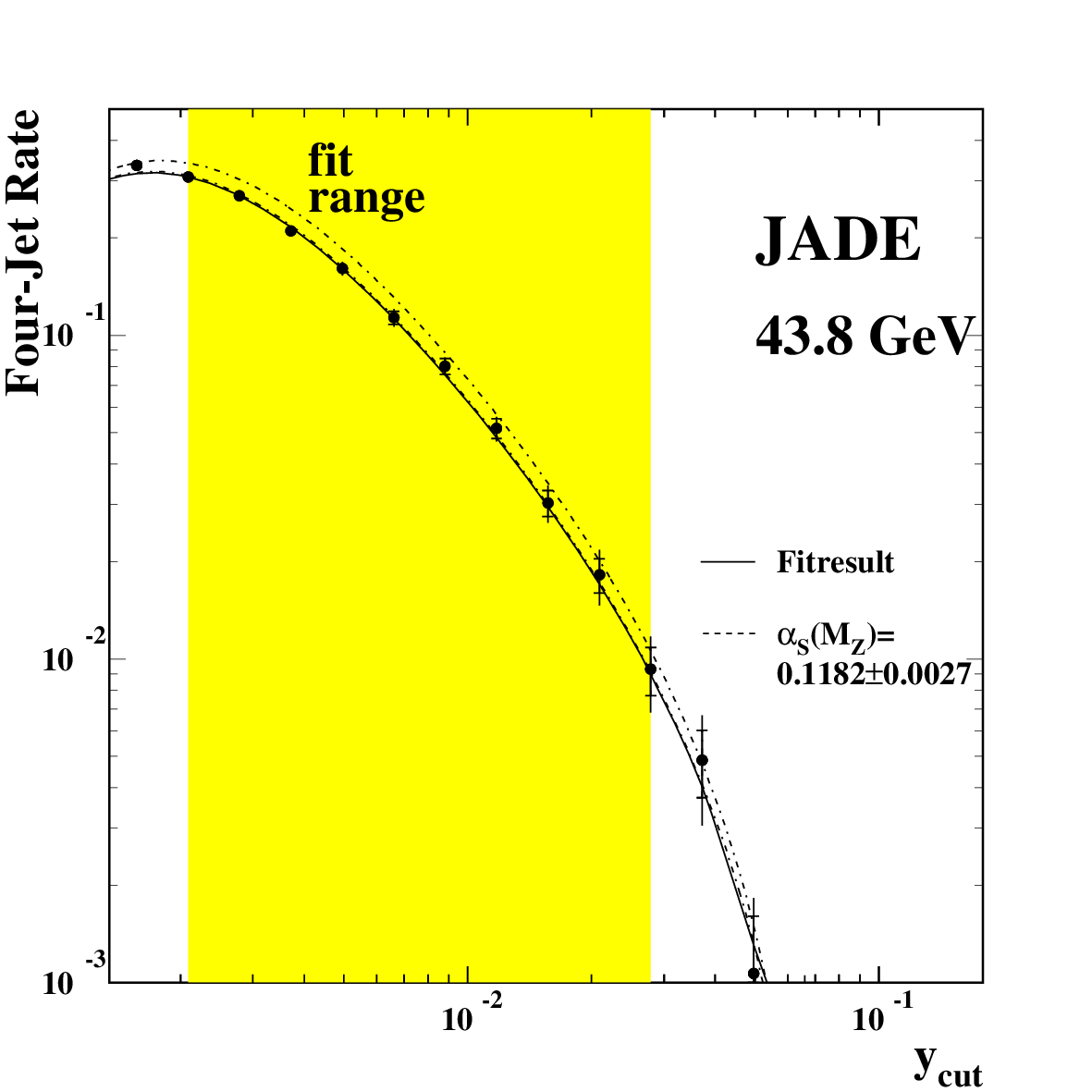} \\
\end{tabular}
\caption{  Same as figure~\ref{fit_plot} for $\rs=34.6$, 35, 38.3 and 43.8~GeV. }
\label{fit_plot2}
\end{figure}
\begin{figure}[htb!]
\begin{center}
\includegraphics[width=1.0\textwidth]{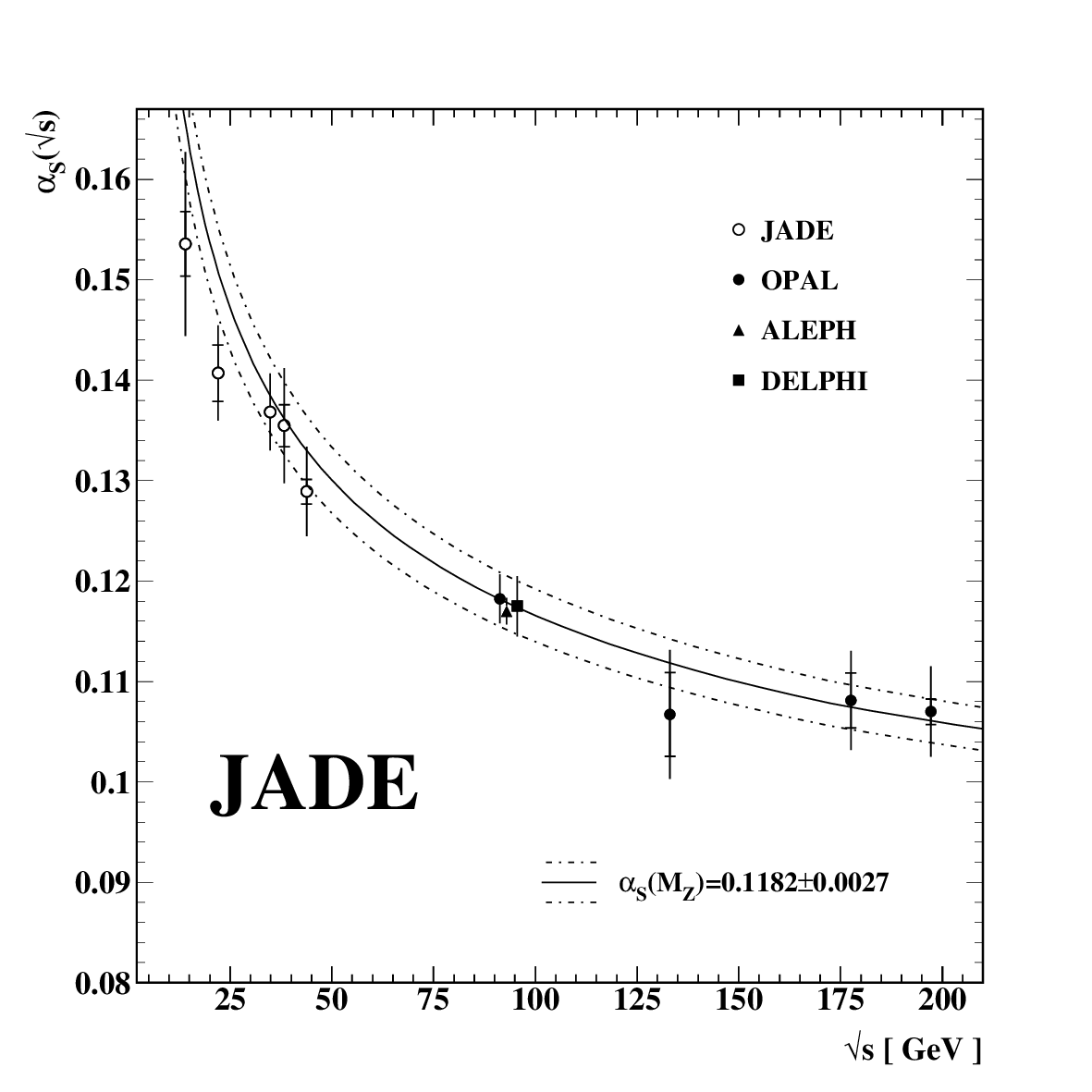}
\end{center}
\caption{The values for \as\ at the various energy points. The errors
  show the statistical (inner part) and the total error.  The full and
  dash-dotted lines indicate the current world average value of
  \asmz~\cite{bethke04} with error.  The results at $\rs=34.6$ and 35~GeV have
  been combined for clarity.  The results from the \Lep\ experiments ALEPH~\cite{aleph249},
  DELPHI~\cite{delphir4} and OPAL~\cite{OPALPN527} are shown as well. }
\label{alphas_fit}
\end{figure}

The differences between the models are considered as a systematic 
uncertainty. \\
A $\chi^{2}$-value at each energy point is calculated using the
following formula:
\be
 \chi^{2} = \sum_{i,j}^{n}(R_{4,i}-R(\as)_{4,i}^{\mathrm{theo}})
            (V(\mathrm{R_{4}})^{-1})_{ij}(R_{4,j}-R(\as)_{4,j}^{\mathrm{theo}})
\ee
where the indices $i$ and $j$ denote the \ycut\ bins in the chosen
fit range and the $R(\as)_{4,i}^{\mathrm{theo}}$ are the predicted
values of the four-jet rate.  
The four-jet rate as a function of \ycut\ is an integrated distribution and
therefore a single event can contribute to several 
\ycut\ bins and neighboring \ycut\ bins are correlated.
The covariance matrix $V_{ij}$ is calculated using 
the matrix $W_{ij}$  
determined from four-jet rate distributions calculated from Monte-Carlo at the 
hadron-level as follows:
\be
W_{ij}=\frac{1}{n-1}\left[\sum_{k=1}^{n} x^{k}_{i}x^{k}_{j}-\frac{1}{n}\sum_{k=1}^{n}x^{k}_{i}
\sum_{k=1}^{n}x^{k}_{j}\right],
\ee
where $x_{i}^{k}$ is the average jet rate for a given \ycut\ bin $i$ and sample $k$, 
and $n$ is the number of subsamples. Subsamples are built by choosing 1000 events
randomly out of the set of all generated Monte 
Carlo events. A single event may be included in several subsamples, 
but the impact on the final covariance matrix 
is expected to be very small and, therefore, is neglected~\cite{bootstrap}.
For every centre-of-mass energy point 1000 subsamples are built. 
The matrix $W_{ij}$ is then used to determine the correlation matrix, 
$\rho_{ij}=W_{ij}/\tilde{\sigma_{i}}\tilde{\sigma_{j}}$, 
with $\tilde{\sigma_{i}}=\sqrt{W_{ii}}$. 
The covariance 
matrix $V_{ij}(R_{4})$ used in
the $\chi^{2}$ fit is then determined using the statistical error 
$\sigma_{i}$ of the data sample at data point $i$ and the correlation matrix 
$\rho_{ij}:V_{ij}(R_{4})=\rho_{ij}\sigma_{i}\sigma_{j}$. \\ 
The $\chi^{2}$ value
is minimized with respect to \as\ for each centre-of-mass energy point
separately. The renormalization scale factor $\xmu$, as discussed 
in section~\ref{theory}, is set to one. \\
The fit ranges are determined by requiring that the hadronization 
corrections  be less than $50\%$ and
the detector corrections be less than $50\%$ in the fit range.
In order to exclude the non-perturbative region we require 
$\sqrt{s\cdot \ycut}$
to be larger than 2~GeV. In the Durham scheme this value corresponds 
to the minimal transverse momentum of the pair of proto-jets
with respect to each other.
The fit range is  $0.0209 < \ycut < 0.0495$ for data taken at 14~GeV,
$0.0088 < \ycut < 0.0495$ for data taken at 22~GeV,
$0.0037 < \ycut < 0.0279$ for data taken at 34.6 and 35~GeV,
$0.0028 < \ycut < 0.0279$ for data taken at 38.3~GeV   
and $0.0021 < \ycut < 0.0279$ for data taken at 43.8~GeV.
In figures~\ref{fit_plot} and~\ref{fit_plot2} the hadron-level
four-jet distributions with the fit results for the six energy points are shown
together with the four-jet rate prediction with \as\ being the current world average.
The results of the fits are summarized
in Table~\ref{fitresults}.  The statistical uncertainties correspond to the
uncertainty from the $\chi^{2}$ minimization, while the systematic errors are
determined as described in section~\ref{systematic}. \\
\begin{table}[]
\begin{center}
\begin{tabular}[tbp]{|c|r|r|r|r|r|r|r|} \hline
\rs\ [GeV] & \asrs & stat. & exp. & hadr. & scale &  \chisqd \\
\hline
$14.00$  & $  0.1536$ &  $  0.0032$ & $  0.0064$ &  $  0.0028$ & $  0.0074$ &  $1.46/3$ \\ 
$22.00$  & $  0.1407$ &  $  0.0028$ & $  0.0034$ &  $  0.0021$ & $  0.0024$ &  $14.22/6$ \\ 
$34.60$  & $  0.1346$ &  $  0.0007$ & $  0.0019$ &  $  0.0031$ & $  0.0011$ &  $17.20/7$ \\ 
$35.00$  & $  0.1391$ &  $  0.0006$ & $  0.0017$ &  $  0.0033$ & $  0.0012$ &  $23.51/7$ \\ 
$38.30$  & $  0.1355$ &  $  0.0021$ & $  0.0042$ &  $  0.0038$ & $  0.0020$ &  $19.78/8$ \\ 
$43.80$  & $  0.1289$ &  $  0.0012$ & $  0.0011$ &  $  0.0038$ & $  0.0019$ &  $4.02/9$ \\ 
\hline
\end{tabular}
\end{center}
\caption{The value of \as\ for each energy point and the statistical,
experimental, hadronization and scale errors.
The last column corresponds to the \chisqd\ value of the fit.}
\label{fitresults}
\end{table}
It is of interest to combine the measurements of \as\ from the
different centre-of-mass energy points in order to determine a single
value of \as\ at a common energy scale. The fit results for \as\ are combined using 
the procedure of Ref.~\cite{pr404}. In brief the method is as 
follows. The set of \as\ measurements to be
combined are first evolved to a common scale, $Q=\mz$, assuming the
validity of QCD and \mz\ being the mass of the \zzero\ vector boson.
The measurements are then combined in a weighted
mean, to minimize the $\chi^{2}$ between the combined value and the
measurements. If the measured values evolved to $Q=\mz$ are denoted
$\asi$, with covariance matrix $V^{\prime}$, the combined value,
\asmz, is given by
\be \asmz=\sum w_{i} \asi \;\;\;\; \mathrm{where}
\;\;\;\;
w_{i}=\frac{\sum_{j}(V^{\prime~-1})_{ij}}{\sum_{j,k}(V^{\prime~-1})_{jk}},
\label{formcomb}
\ee
where $i$ and $j$ denote the six individual results.  The difficulty
resides in making a reliable estimate of $V^{\prime}$ in the presence
of dominant and highly correlated systematic errors.  Small
uncertainties in the estimation of these correlations can cause
undesirable features such as negative weights.  For this reason only
experimental systematic errors assumed to be partially correlated
between measurements contribute to the off-diagonal
elements of the covariance matrix:
$V^{\prime}_{ij}=\min(\sigma^2_{\mathrm{exp},i},\sigma^2_{\mathrm{exp},j})$.  All error
contributions (statistical, experimental, hadronization and scale
uncertainty) contribute to the diagonal elements only.  The
hadronization and scale uncertainties are computed by combining the
\as\ values obtained with the alternative hadronization models, and
from the upper and lower theoretical errors, using the weights derived
from the covariance matrix $V^{\prime}$.  \\
The fit result from the 14~GeV data has large
experimental and theoretical uncertainties.
We therefore choose to not include this result in the combination.  
The combination using all results for $\rs\ge 22$~GeV is
\be
 \result\;,
\label{mainresult}
\ee
consistent with the world average value of $\asmz=0.1182\pm0.0027$
~\cite{bethke04}.  The weights $w_{i}$ as described
in Eq.~\ref{formcomb} are 
0.15 for 22~GeV, 0.29 for
34.6~GeV, 0.29 for 35~GeV, 0.06 for 38.3~GeV and 0.21 for 44~GeV.  
The results at each energy point are shown in figure~\ref{alphas_fit}
and compared with the predicted running of \as\ based QCD and on the world
average value of \asmz.  For clarity the values at 34.6 and 35~GeV
have been combined at their luminosity weighted average energy
$\rs=34.8$~GeV using the combination procedure described above.
The combined value is \resultthirty.
The results of ALEPH~\cite{aleph249}, DELPHI~\cite{delphir4} and OPAL~\cite{OPALPN527}
 are shown as well.
\subsection{Renormalization Scale Dependence of \boldmath{\as}}
\label{r4scalefree}
\begin{figure}[htb!]
\begin{center}
\begin{tabular}[tbp]{cc}
\includegraphics[width=0.4\textwidth]{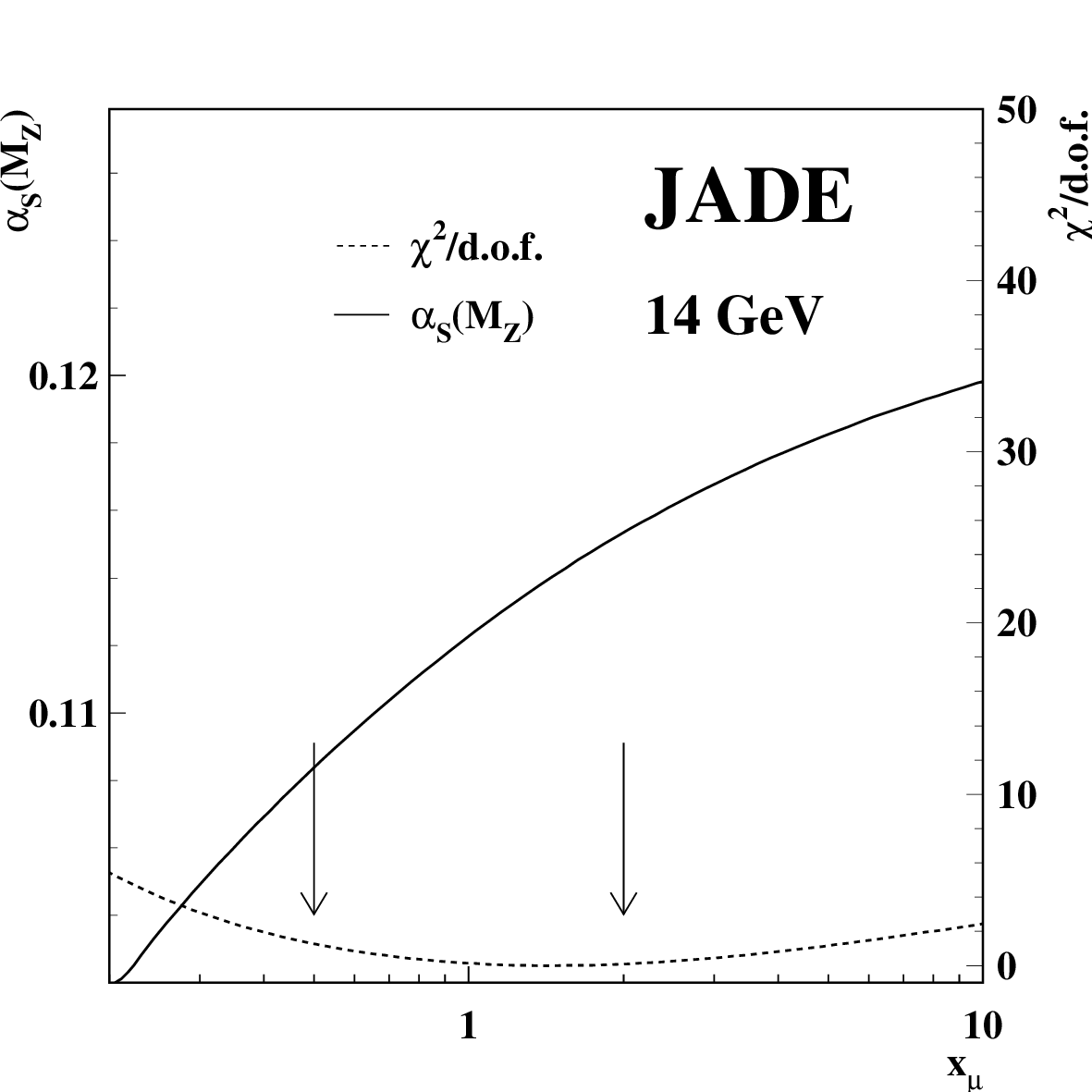} &
\includegraphics[width=0.4\textwidth]{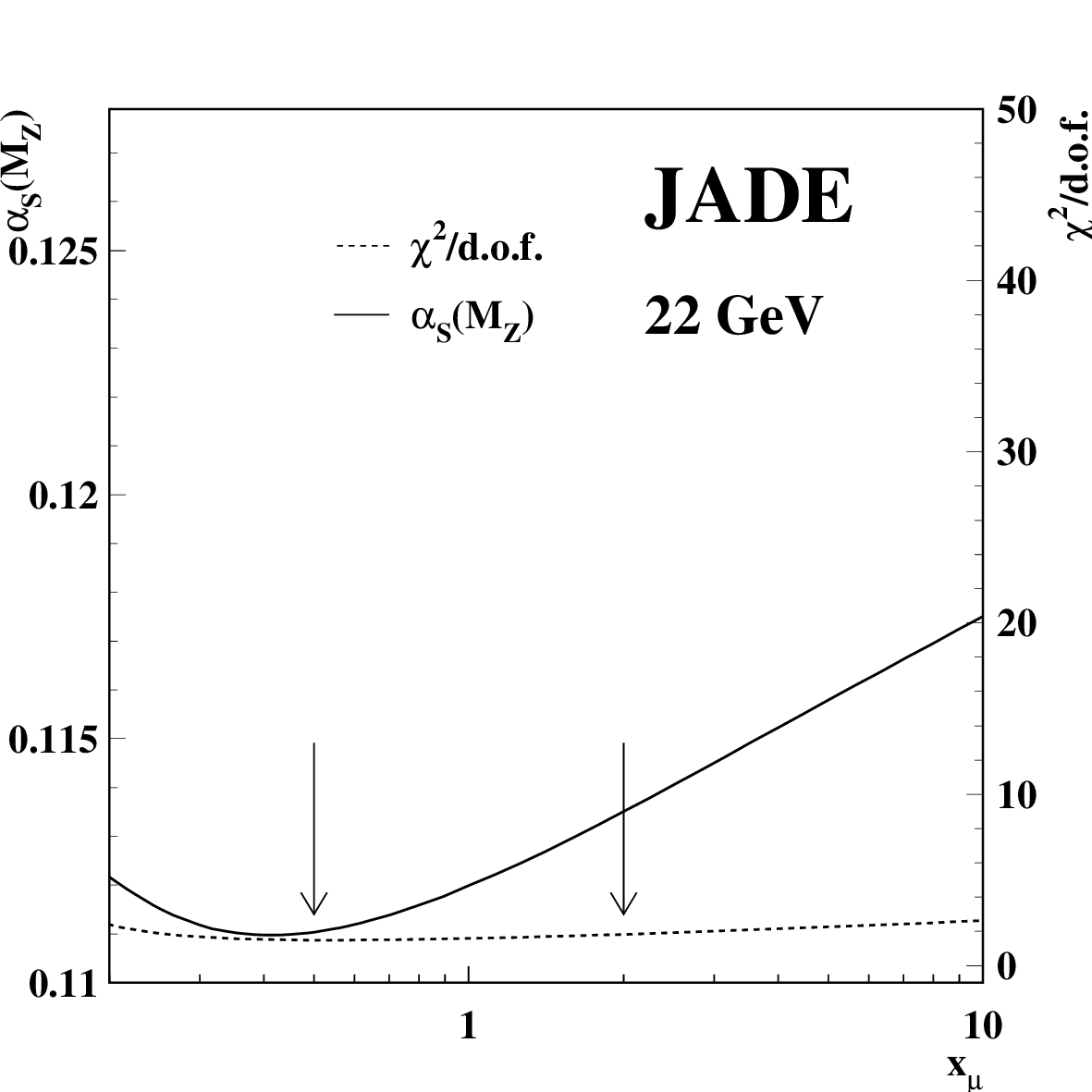} \\
\includegraphics[width=0.4\textwidth]{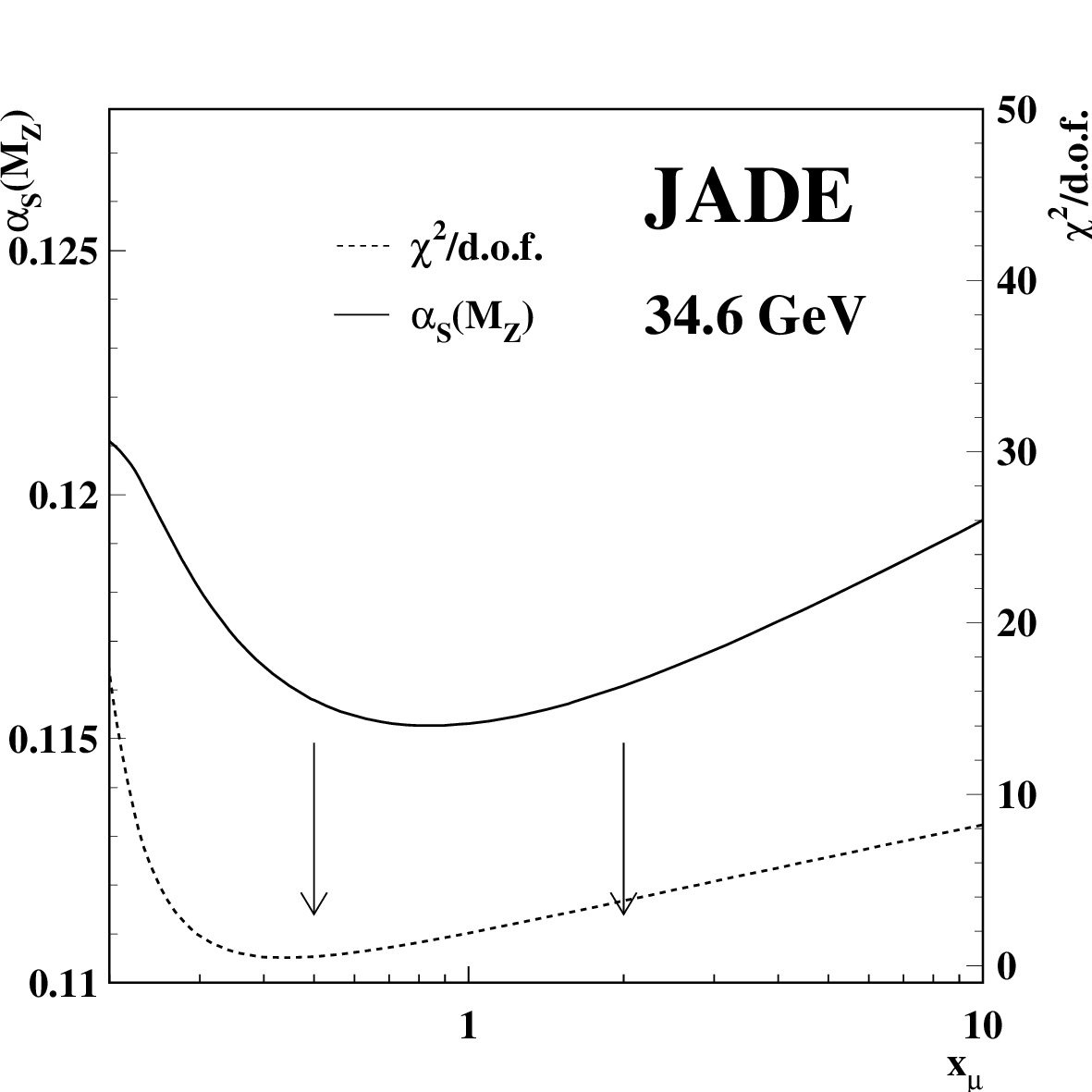} &
\includegraphics[width=0.4\textwidth]{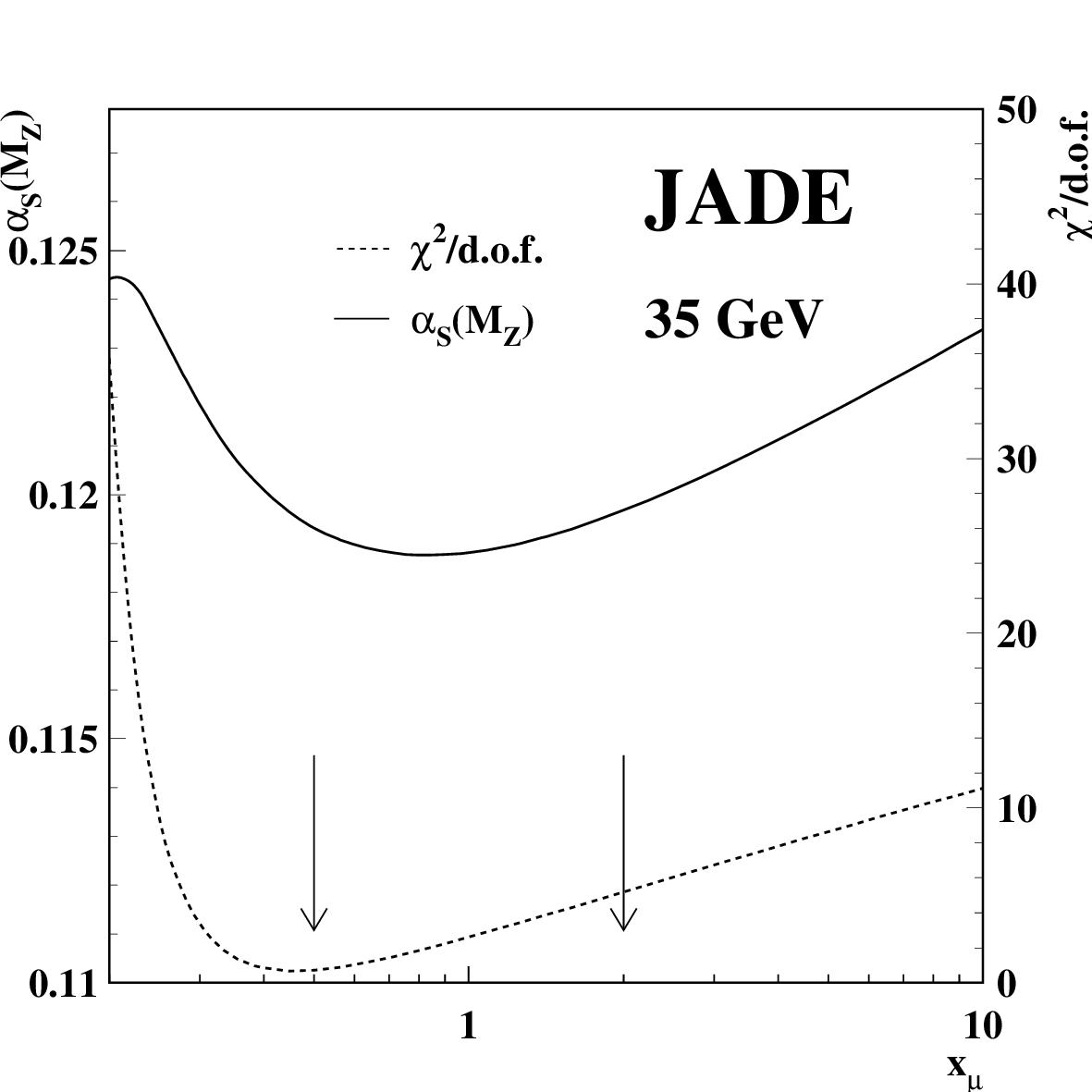} \\
\includegraphics[width=0.4\textwidth]{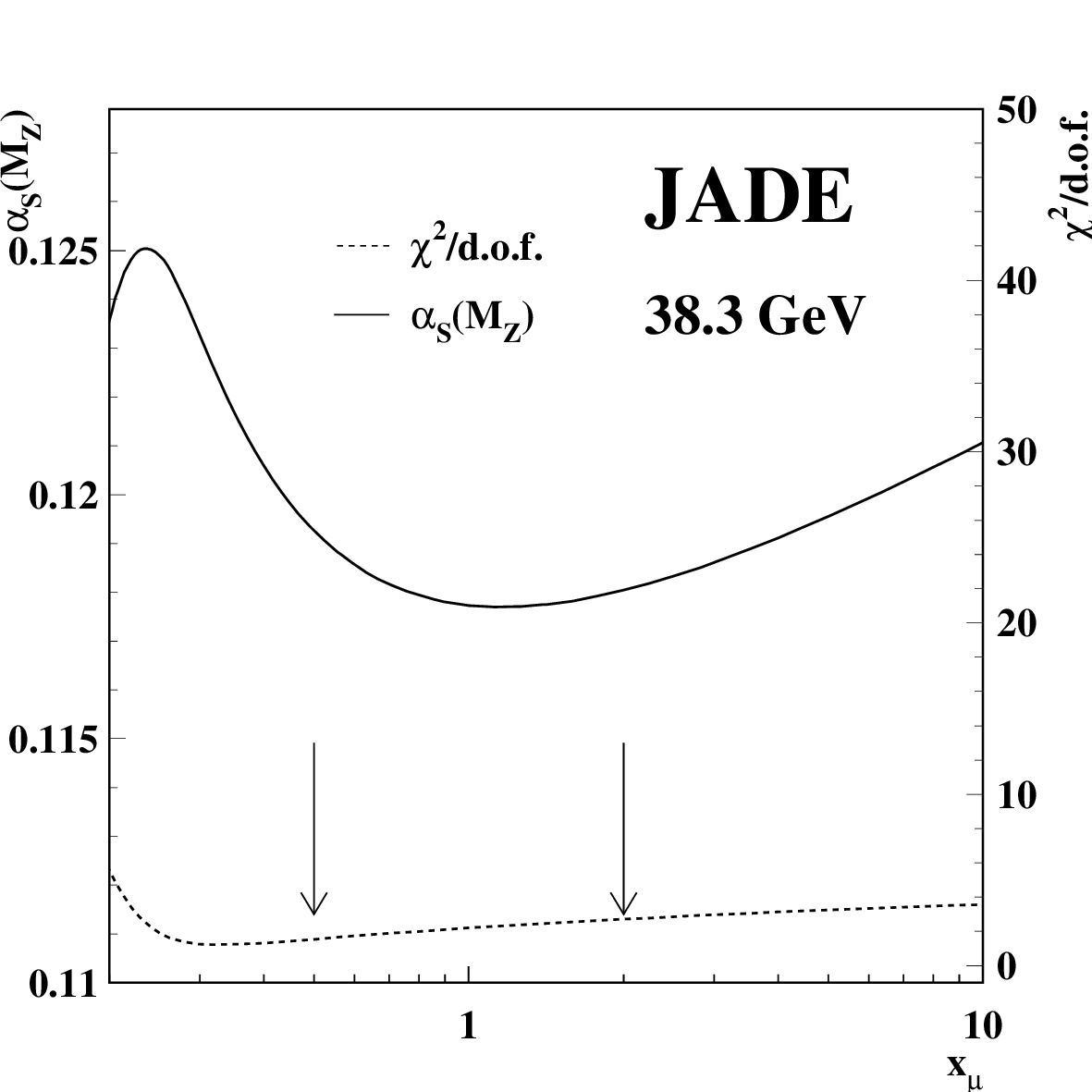} &
\includegraphics[width=0.4\textwidth]{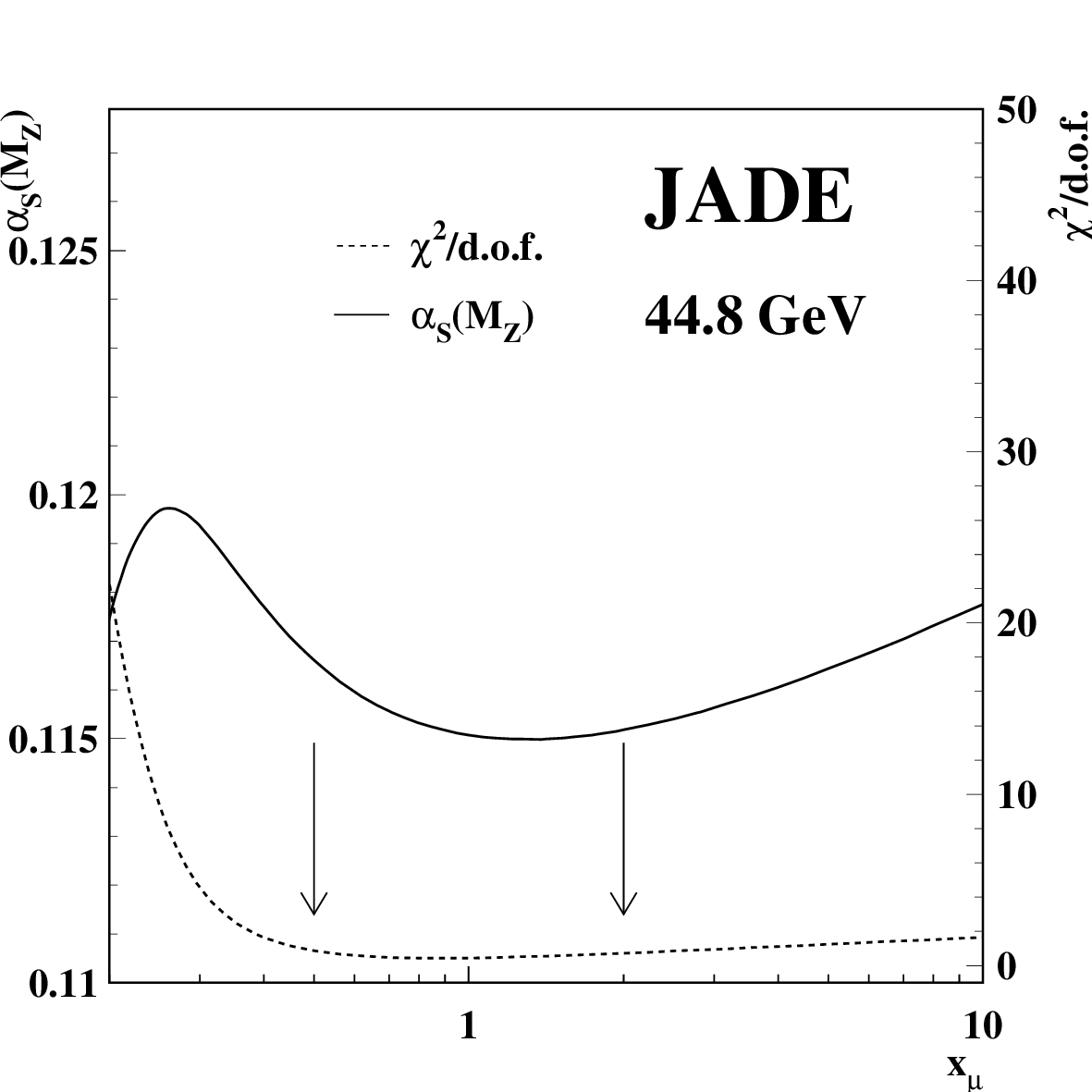} \\
\end{tabular}
\end{center}
\caption{The result of \asmz\ and the \chisqd\ of the 
fit to the four-jet rate as a function of the renormalization 
scale \xmu\ for $\rs=14$~GeV to 43.8~GeV.
The arrows indicate the variation of the renormalization
scale factor used for the
determination of the systematic uncertainties.
}
\label{xmuopt}
\end{figure}

For the fits in Section~\ref{fitprocedure} the renormalization 
scale factor is set to the natural choice $\xmu=1$, where
the energy dependence of Eq.~\ref{NLLA} comes only from the
running of the coupling \as. However, different schemes for 
the determination of the renormalization scale factor are 
proposed.
As a cross check of our default result we investigate 
in this section two more choices 
of the renormalization scale factor.  \\
In the optimized renormalization scheme~\cite{xmuopt} the 
minimization is performed for all energy points 
with \as\ and \xmu\ treated as free parameters. The systematic
uncertainties are determined using the optimized renormalization
scale determined in the default fit.
The value of \as, the optimized scale \xmuopt\ and their correlation are summarized 
in Table ~\ref{fitresultsxmufree}.
\begin{table}[htb!]
\begin{center}
\begin{tabular}[tbp]{|c|r|r|r|r|c|c|r|} \hline
\rs\ [GeV] & \asrs & stat. & exp. & hadr. &  \xmuopt & Corr. & \chisqd \\
\hline
$14.00$  & $  0.1572$ &  $  0.0033$ & $  0.0065$ &  $  0.0031$ & $ 1.49  \pm 0.50 $ &  0.66 & $0.01/2$ \\ 
$22.00$  & $  0.1393$ &  $  0.0027$ & $  0.0033$ &  $  0.0019$ & $ 0.55  \pm 0.31 $ &  0.28 & $13.38/5$ \\ 
$34.60$  & $  0.1357$ &  $  0.0007$ & $  0.0019$ &  $  0.0030$ & $ 0.44  \pm 0.07 $ &  -0.68 & $4.09/6$ \\ 
$35.00$  & $  0.1401$ &  $  0.0006$ & $  0.0017$ &  $  0.0031$ & $ 0.46  \pm 0.06 $ &  -0.67 & $6.04/6$ \\ 
$38.30$  & $  0.1416$ &  $  0.0024$ & $  0.0049$ &  $  0.0039$ & $ 0.33  \pm 0.05 $ &  -0.68 & $10.98/7$ \\ 
$43.80$  & $  0.1291$ &  $  0.0012$ & $  0.0011$ &  $  0.0037$ & $ 0.87  \pm 0.33 $ &  -0.44 & $3.90/8$ \\ 
\hline
\end{tabular}
\end{center}
\caption{The value of \as\ and the statistical,
experimental, hadronization, renormalization scale factor 
\xmuopt, the correlation between \as\ and the 
renormalization scale factor and the \chisqd\ value of the fit
for all energy points with the minimization performed using the optimized 
renormalization scheme.}
\label{fitresultsxmufree}
\end{table}
The variation of \chisqd\ as a function of the scale \xmu\ is 
shown in Figure~\ref{xmuopt}. 
The combination of all energy points at $\rs\ge 22$~GeV 
using the method described 
in Section~\ref{fitprocedure}~\footnote{In this case the statistical, 
experimental and hadronization uncertainty only
contribute to the diagonal elements of the covariance matrix
$V^{\prime}$.} yields
\be
\resultxmu.
\ee
The values for \xmuopt\ for centre-of-mass energies
of 14, 22, 34.6, 35 and 43.8~GeV are within their
statistical uncertainties covered by the 
systematic variation ($0.5 < \xmu < 2.0$) of 
the default fit. \\
The second choice for the determination of the renormalization scale factor 
followed approximately the approach of ``minimal sensitivity'' (PMS)
suggested by~\cite{scaleasmin}. 
The renormalization scale factor \xmumin\ is specified by the point with \as\ 
having the least sensitivity to the renormalization scale factor \xmu. The 
variation  of \as\ as a function of \xmu\ is  
shown in Figure~\ref{xmuopt}. The renormalization scale factor \xmumin\ 
is determined by a fit to the \as\ variation with respect to \xmu\ around
the minimum. The determination of \as\ is then repeated 
with \xmu\ set to \xmumin. At 14~GeV the variation of \as\ with respect
to the renormalization scale factor has no minimum and therefore
no fit is performed.
The renormalization scale factor \xmumin\ as well as the 
result of the fit are summarized in Table~\ref{fitresultsxmin}.
The systematic uncertainties are determined using \xmumin\ determined 
in the default fit.
\begin{table}[]
\begin{center}
\begin{tabular}[tbp]{|c|r|r|c|r|r|r|r|} \hline
\rs\ [GeV] & \asrs & stat. & exp. & hadr. & \xmumin &  \chisqd \\
\hline
$22.00$  & $  0.1391$ &  $  0.0027$ & $  0.0034$ &  $  0.0018$ & $0.42$ & $13.61/6$ \\ 
$34.60$  & $  0.1345$ &  $  0.0007$ & $  0.0019$ &  $  0.0031$ & $0.92$ & $15.38/7$ \\ 
$35.00$  & $  0.1391$ &  $  0.0006$ & $  0.0017$ &  $  0.0033$ & $0.92$ & $20.78/7$ \\ 
$38.30$  & $  0.1354$ &  $  0.0021$ & $  0.0042$ &  $  0.0038$ & $1.15$ & $20.80/8$ \\ 
$43.80$  & $  0.1288$ &  $  0.0012$ & $  0.0011$ &  $  0.0038$ & $1.22$ & $4.50/9$ \\ 
\hline
\end{tabular}
\end{center}
\caption{The value of \as\ and the statistical,
experimental, hadronization, the renormalization 
scale scale factor \xmumin\ and the \chisqd\ value of the fit 
for the energy points between 22 and 43.8~GeV with \as\ having 
the least sensitivity to the renormalization
scale \xmu.}
\label{fitresultsxmin}
\end{table}
The combination of all energy points above $\rs\ge 22$~GeV 
using the method described  in 
Section~\ref{fitprocedure}~\footnotemark[\value{footnote}] yields
\be
\resultxmin,
\ee
consistent with Eq.~\ref{mainresult}.
The local minimum of \as\ as a function of \xmu, is very close
the natural scale $\xmu=1$ 
leading to a 
fitted value of \as\ similar to the default value Eq.~\ref{mainresult}. \\
The choice of the renormalization scale factor \xmumin\ and \xmuopt\ returns
a value of \as\ which is within the variation of the systematic 
uncertainty due to missing higher order terms.
We therefore conclude that the evaluation of theoretical
uncertainties using the commonly used standard method of setting 
\xmu\ to $0.5$ and $2.$ safely includes alternative methods
of the choice of the renormalization scale factor.
\section{Summary}
In this paper we present measurements of the strong
coupling from the four-jet rate at centre-of-mass energies between 14
and 44~GeV using data of the JADE experiment.  The predictions of the
\py, \hw\ and \ar\ Monte Carlo models tuned by OPAL to LEP~1 data are
found to be in agreement with the measured distributions.\\
From a fit of QCD NLO predictions combined with resummed NLLA
calculations with \xmu=1 to the four-jet rate corrected 
for experimental and hadronization effects we have determined 
the strong coupling \as. In addition we investigated two more
choices for the determination of the renormalization scale
and found the results to be consistent. 
The value of \asmz\ is determined to be \restot.
The natural choice of the renormalization scale $\xmu=1$
is close to the renormalization scale factor with \as\ having
the least sensitivity to the renormalization scale factor, 
$\frac{\mathrm{d}\as}{\mathrm{d}\xmu}=0$. 
Therefore the theoretical uncertainty determined by setting the scale \xmu\ to
\xmu=0.5  and \xmu=2.0 yields smaller theoretical uncertainties 
than e.g. for fits to event shape observables~\cite{pedrophd}.
This is also true for measurements of \as\ performed at LEP 
energies~\cite{OPALPN527}.

\clearpage
\begin{table}[]
\begin{center}
\begin{tabular}[tbp]{|c|c|c|c|} \hline
$\log_{10}(\ycut)$ & $R_{4}$(14~GeV) & $R_{4}$(22~GeV) & $R_{4}$(34.6~GeV) \\
\hline
$ -4.68 $ &&  & $ 0.004  \pm 0.001 \pm 0.005 $  \\ 
$ -4.55 $ && $ 0.008  \pm 0.002 \pm 0.014 $ & $ 0.002  \pm 0.000 \pm 0.002 $  \\ 
$ -4.43 $ && $ 0.002  \pm 0.001 \pm 0.003 $ & $ 0.002  \pm 0.000 \pm 0.002 $  \\ 
$ -4.30 $ && $ 0.002  \pm 0.001 \pm 0.002 $ & $ 0.002  \pm 0.000 \pm 0.001 $  \\ 
$ -4.18 $ && $ 0.002  \pm 0.001 \pm 0.001 $ & $ 0.002  \pm 0.000 \pm 0.001 $  \\ 
$ -4.05 $ && $ 0.001  \pm 0.001 \pm 0.001 $ & $ 0.002  \pm 0.000 \pm 0.001 $  \\ 
$ -3.93 $ & $ 0.001  \pm 0.001 \pm 0.001 $ & $ 0.002  \pm 0.001 \pm 0.000 $ & $ 0.003  \pm 0.001 \pm 0.001 $  \\ 
$ -3.81 $ & $ 0.001  \pm 0.001 \pm 0.002 $ & $ 0.003  \pm 0.002 \pm 0.001 $ & $ 0.004  \pm 0.001 \pm 0.001 $  \\ 
$ -3.68 $ & $ 0.003  \pm 0.001 \pm 0.003 $ & $ 0.004  \pm 0.002 \pm 0.002 $ & $ 0.008  \pm 0.001 \pm 0.001 $  \\ 
$ -3.56 $ & $ 0.003  \pm 0.001 \pm 0.003 $ & $ 0.007  \pm 0.002 \pm 0.002 $ & $ 0.014  \pm 0.001 \pm 0.001 $  \\ 
$ -3.43 $ & $ 0.005  \pm 0.002 \pm 0.003 $ & $ 0.009  \pm 0.003 \pm 0.002 $ & $ 0.027  \pm 0.001 \pm 0.002 $  \\ 
$ -3.31 $ & $ 0.008  \pm 0.002 \pm 0.004 $ & $ 0.015  \pm 0.003 \pm 0.003 $ & $ 0.054  \pm 0.002 \pm 0.003 $  \\ 
$ -3.18 $ & $ 0.016  \pm 0.003 \pm 0.006 $ & $ 0.030  \pm 0.005 \pm 0.004 $ & $ 0.099  \pm 0.003 \pm 0.004 $  \\ 
$ -3.06 $ & $ 0.028  \pm 0.004 \pm 0.009 $ & $ 0.063  \pm 0.007 \pm 0.006 $ & $ 0.169  \pm 0.003 \pm 0.005 $  \\ 
$ -2.93 $ & $ 0.052  \pm 0.005 \pm 0.011 $ & $ 0.104  \pm 0.008 \pm 0.012 $ & $ 0.252  \pm 0.004 \pm 0.004 $  \\ 
$ -2.81 $ & $ 0.090  \pm 0.006 \pm 0.012 $ & $ 0.185  \pm 0.011 \pm 0.012 $ & $ 0.316  \pm 0.004 \pm 0.006 $  \\ 
$ -2.68 $ & $ 0.155  \pm 0.008 \pm 0.017 $ & $ 0.269  \pm 0.012 \pm 0.007 $ & $ 0.341  \pm 0.004 \pm 0.003 $  \\ 
$ -2.56 $ & $ 0.229  \pm 0.009 \pm 0.015 $ & $ 0.345  \pm 0.013 \pm 0.021 $ & $ 0.326  \pm 0.004 \pm 0.003 $  \\ 
$ -2.43 $ & $ 0.327  \pm 0.011 \pm 0.017 $ & $ 0.379  \pm 0.013 \pm 0.016 $ & $ 0.273  \pm 0.004 \pm 0.009 $  \\ 
$ -2.31 $ & $ 0.391  \pm 0.011 \pm 0.010 $ & $ 0.361  \pm 0.013 \pm 0.031 $ & $ 0.211  \pm 0.004 \pm 0.008 $  \\ 
$ -2.18 $ & $ 0.405  \pm 0.011 \pm 0.011 $ & $ 0.265  \pm 0.012 \pm 0.007 $ & $ 0.156  \pm 0.003 \pm 0.007 $  \\ 
$ -2.06 $ & $ 0.375  \pm 0.011 \pm 0.015 $ & $ 0.182  \pm 0.011 \pm 0.013 $ & $ 0.106  \pm 0.003 \pm 0.005 $  \\ 
$ -1.93 $ & $ 0.291  \pm 0.010 \pm 0.009 $ & $ 0.120  \pm 0.009 \pm 0.013 $ & $ 0.069  \pm 0.002 \pm 0.004 $  \\ 
$ -1.80 $ & $ 0.189  \pm 0.009 \pm 0.015 $ & $ 0.084  \pm 0.008 \pm 0.013 $ & $ 0.040  \pm 0.002 \pm 0.002 $  \\ 
$ -1.68 $ & $ 0.099  \pm 0.007 \pm 0.012 $ & $ 0.041  \pm 0.005 \pm 0.012 $ & $ 0.023  \pm 0.001 \pm 0.002 $  \\ 
$ -1.55 $ & $ 0.043  \pm 0.004 \pm 0.007 $ & $ 0.026  \pm 0.004 \pm 0.003 $ & $ 0.012  \pm 0.001 \pm 0.001 $  \\ 
$ -1.43 $ & $ 0.014  \pm 0.003 \pm 0.006 $ & $ 0.011  \pm 0.003 \pm 0.003 $ & $ 0.005  \pm 0.001 \pm 0.001 $  \\ 
$ -1.30 $ & $ 0.003  \pm 0.001 \pm 0.002 $ & $ 0.002  \pm 0.001 \pm 0.001 $ & $ 0.002  \pm 0.000 \pm 0.000 $  \\ 
$ -1.18 $ & $ -0.001  \pm 0.001 \pm 0.001 $ &   &  \\ 
\hline
\end{tabular}
\end{center}
\caption{
Hadron-level value of the four-jet fraction
using the Durham algorithm at 14, 22 and 34.6 GeV.
In all cases the first quoted error indicates the statistical
error while the second quoted error corresponds to the
total experimental uncertainty. Uncertainties consistent with zero
indicate that the corresponding value is smaller than
the precision shown in the table.}
\label{hadron_tab_1}
\end{table}
\begin{table}[]
\begin{center}
\begin{tabular}[tbp]{|c|c|c|c|} \hline
$\log_{10}(\ycut)$ & $R_{4}$(35~GeV) & $R_{4}$(38.3~GeV) & $R_{4}$(43.8~GeV) \\
\hline
$ -4.80 $ & $ 0.001  \pm 0.000 \pm 0.001 $ &  &  \\ 
$ -4.68 $ & $ 0.003  \pm 0.000 \pm 0.004 $ & $ 0.001  \pm 0.001 \pm 0.001 $ &  \\ 
$ -4.55 $ & $ 0.002  \pm 0.000 \pm 0.002 $ & $ 0.001  \pm 0.001 \pm 0.001 $ & $ 0.011  \pm 0.002 \pm 0.014 $  \\ 
$ -4.43 $ & $ 0.002  \pm 0.000 \pm 0.002 $ & $ 0.004  \pm 0.002 \pm 0.004 $ & $ 0.004  \pm 0.001 \pm 0.004 $  \\ 
$ -4.30 $ & $ 0.002  \pm 0.000 \pm 0.002 $ & $ 0.005  \pm 0.002 \pm 0.004 $ & $ 0.004  \pm 0.001 \pm 0.004 $  \\ 
$ -4.18 $ & $ 0.003  \pm 0.000 \pm 0.002 $ & $ 0.004  \pm 0.002 \pm 0.002 $ & $ 0.004  \pm 0.001 \pm 0.003 $  \\ 
$ -4.05 $ & $ 0.002  \pm 0.000 \pm 0.001 $ & $ 0.005  \pm 0.002 \pm 0.002 $ & $ 0.004  \pm 0.001 \pm 0.002 $  \\ 
$ -3.93 $ & $ 0.003  \pm 0.000 \pm 0.001 $ & $ 0.005  \pm 0.002 \pm 0.002 $ & $ 0.004  \pm 0.001 \pm 0.001 $  \\ 
$ -3.81 $ & $ 0.005  \pm 0.001 \pm 0.001 $ & $ 0.007  \pm 0.002 \pm 0.002 $ & $ 0.006  \pm 0.001 \pm 0.001 $  \\ 
$ -3.68 $ & $ 0.007  \pm 0.001 \pm 0.000 $ & $ 0.011  \pm 0.003 \pm 0.002 $ & $ 0.014  \pm 0.002 \pm 0.001 $  \\ 
$ -3.56 $ & $ 0.014  \pm 0.001 \pm 0.002 $ & $ 0.019  \pm 0.004 \pm 0.004 $ & $ 0.027  \pm 0.003 \pm 0.002 $  \\ 
$ -3.43 $ & $ 0.027  \pm 0.001 \pm 0.003 $ & $ 0.032  \pm 0.005 \pm 0.005 $ & $ 0.055  \pm 0.004 \pm 0.003 $  \\ 
$ -3.31 $ & $ 0.054  \pm 0.002 \pm 0.004 $ & $ 0.068  \pm 0.007 \pm 0.006 $ & $ 0.105  \pm 0.005 \pm 0.007 $  \\ 
$ -3.18 $ & $ 0.100  \pm 0.002 \pm 0.006 $ & $ 0.118  \pm 0.009 \pm 0.015 $ & $ 0.181  \pm 0.006 \pm 0.006 $  \\ 
$ -3.06 $ & $ 0.171  \pm 0.003 \pm 0.004 $ & $ 0.191  \pm 0.011 \pm 0.008 $ & $ 0.265  \pm 0.007 \pm 0.006 $  \\ 
$ -2.93 $ & $ 0.254  \pm 0.003 \pm 0.011 $ & $ 0.267  \pm 0.012 \pm 0.013 $ & $ 0.323  \pm 0.008 \pm 0.014 $  \\ 
$ -2.81 $ & $ 0.316  \pm 0.004 \pm 0.011 $ & $ 0.325  \pm 0.013 \pm 0.018 $ & $ 0.335  \pm 0.008 \pm 0.007 $  \\ 
$ -2.68 $ & $ 0.344  \pm 0.004 \pm 0.004 $ & $ 0.328  \pm 0.013 \pm 0.013 $ & $ 0.308  \pm 0.008 \pm 0.006 $  \\ 
$ -2.56 $ & $ 0.331  \pm 0.004 \pm 0.004 $ & $ 0.297  \pm 0.013 \pm 0.013 $ & $ 0.270  \pm 0.007 \pm 0.006 $  \\ 
$ -2.43 $ & $ 0.289  \pm 0.003 \pm 0.009 $ & $ 0.253  \pm 0.012 \pm 0.014 $ & $ 0.210  \pm 0.007 \pm 0.005 $  \\ 
$ -2.31 $ & $ 0.231  \pm 0.003 \pm 0.006 $ & $ 0.196  \pm 0.011 \pm 0.005 $ & $ 0.161  \pm 0.006 \pm 0.005 $  \\ 
$ -2.18 $ & $ 0.168  \pm 0.003 \pm 0.003 $ & $ 0.150  \pm 0.010 \pm 0.009 $ & $ 0.113  \pm 0.005 \pm 0.005 $  \\ 
$ -2.06 $ & $ 0.113  \pm 0.002 \pm 0.003 $ & $ 0.117  \pm 0.009 \pm 0.006 $ & $ 0.080  \pm 0.004 \pm 0.002 $  \\ 
$ -1.93 $ & $ 0.074  \pm 0.002 \pm 0.002 $ & $ 0.089  \pm 0.008 \pm 0.006 $ & $ 0.052  \pm 0.004 \pm 0.003 $  \\ 
$ -1.80 $ & $ 0.044  \pm 0.002 \pm 0.002 $ & $ 0.058  \pm 0.006 \pm 0.007 $ & $ 0.030  \pm 0.003 \pm 0.003 $  \\ 
$ -1.68 $ & $ 0.025  \pm 0.001 \pm 0.002 $ & $ 0.035  \pm 0.005 \pm 0.004 $ & $ 0.018  \pm 0.002 \pm 0.003 $  \\ 
$ -1.55 $ & $ 0.012  \pm 0.001 \pm 0.001 $ & $ 0.018  \pm 0.004 \pm 0.002 $ & $ 0.009  \pm 0.002 \pm 0.002 $  \\ 
$ -1.43 $ & $ 0.007  \pm 0.001 \pm 0.001 $ & $ 0.008  \pm 0.003 \pm 0.002 $ & $ 0.005  \pm 0.001 \pm 0.001 $  \\ 
$ -1.30 $ & $ 0.004  \pm 0.001 \pm 0.001 $ & $ 0.003  \pm 0.001 \pm 0.004 $ & $ 0.001  \pm 0.001 \pm 0.001 $  \\ 
$ -1.18 $ &  & $ 0.001  \pm 0.001 \pm 0.001 $ & $ 0.001  \pm 0.000 \pm 0.000 $  \\ 
\hline
\end{tabular}
\end{center}
\caption{
Hadron-level value of the four-jet fraction
using the Durham algorithm at 35, 38.3 and 43.8 GeV.
In all cases the first quoted error indicates the statistical
error while the second quoted error corresponds to the
total experimental uncertainty. Uncertainties consistent with zero
indicate that the corresponding value is smaller than
the precision shown in the table.}
\label{hadron_tab_2}
\end{table}
\clearpage

\end{document}